\documentclass[11pt,a4paper]{article}
\pdfoutput=1
\usepackage{jheppub}
\usepackage{amsfonts, amssymb, amsmath}
\usepackage{mathrsfs}
\usepackage{graphicx}
\usepackage[utf8]{inputenc}
\usepackage[russian,english]{babel}

\usepackage{color}
\definecolor{dark-gray}{gray}{0.20}
\definecolor{gray}{gray}{0.30}
\definecolor{light-gray}{gray}{0.80}
\definecolor{dark-red}{rgb}{0.7,0,0}
\definecolor{dark-green}{rgb}{0.1,0.4,0}
\definecolor{dark-blue}{rgb}{0.3,0.3,0.7}
\definecolor{light-blue}{rgb}{0.8,0.8,1}
\definecolor{blue}{rgb}{0,0,1}
\definecolor{red}{rgb}{1,0,0}
\definecolor{green}{rgb}{0,1,0}

\usepackage{hyperref}
\hypersetup{
	colorlinks=true,
	linkcolor=blue,
	citecolor=blue,
	urlcolor=blue,
	linktoc=page
}

\def\cB{{\cal B}}

\def\cF{{\cal F}}

\def\cN{{\cal N}}

\def\sl{\frak{sl}}

\newcommand{\be}{\begin{equation}}
\newcommand{\ee}{\end{equation}}
\newcommand{\bea}{\begin{eqnarray}}
\newcommand{\eea}{\end{eqnarray}}

\title{Spindle black holes and theories of class $\mathcal{F}$}

\author[a,b]{Kiril Hristov}
\author[c]{and Minwoo Suh}

\affiliation[a]{Faculty of Physics, Sofia University, \\ J. Bourchier Blvd. 5, 1164 Sofia, Bulgaria}
\affiliation[b]{INRNE, Bulgarian Academy of Sciences, \\ Tsarigradsko Chaussee 72, 1784 Sofia, Bulgaria}
\affiliation[c]{School of General Education, Kumoh National Institute of Technology, \\ Gumi, 39177, Korea}

\emailAdd{khristov@phys.uni-sofia.bg}
\emailAdd{minwoosuh1@gmail.com}

\abstract{We construct solutions in the form of $AdS_2\times\Sigma$, where $\Sigma$ denotes a spindle geometry, within a particular 4d $\mathcal{N}=2$ gauged supergravity with vector multiplets and a charged hypermultiplet stemming from a consistent truncation of 6d matter coupled $F(4)$ gauged supergravity on a Riemann surface. We also compute the Bekenstein-Hawking entropy for the presumed asymptotically $AdS_4$ black hole solutions, providing a dual prediction for the spindle index of 3d superconformal field theories falling under the class $\mathcal{F}$. Notably, the gravitational block description of the solutions accurately replicates the entropy and the values of scalar fields at the two orbifold points of the spindle, validating our findings.}
\date{\today}
\begin{document}
\maketitle


\section{Introduction}
\label{sec:1}

The twisted compactification stands out as a highly potent tool for generating novel solutions or theories within the realms of supergravity, \cite{Maldacena:2000mw}, and field theory, \cite{Gaiotto:2009we}. A prime illustration of its efficacy lies in the derivation of four-dimensional class $\mathcal{S}$ theories from six-dimensional $(2,0)$ theories, \cite{Gaiotto:2009we}, along with their corresponding gravity duals, \cite{Maldacena:2000mw, Bah:2011vv, Bah:2012dg}. Similarly, this approach allows for the construction of three-dimensional superconformal field theories (SCFTs) through the compactification of five-dimensional theories, called the theories of class $\mathcal{F}$, \cite{Bah:2018lyv}.

A notable class of five-dimensional SCFTs emerge from the decoupling limit of $N$ D4-branes accompanied by $N_f<8$ D8-branes and an $O8$-plane in the background. These theories retain eight supercharges and boast a flavor symmetry of $SU(2)\times{E}_{N_f+1}$, \cite{Seiberg:1996bd}. According to the AdS/CFT correspondence, \cite{Maldacena:1997re}, the gravitational duals of these theories manifest as $AdS_6\times{HS}^4$ solutions within massive type IIA supergravity, \cite{Brandhuber:1999np}, where $HS^4$ denotes a four-dimensional hemisphere. Alternatively, these solutions can be derived by uplifting the supersymmetric $AdS_6$ critical point of $F(4)$ gauged supergravity in six dimensions, \cite{Romans:1985tw}, to massive type IIA supergravity, \cite{Cvetic:1999un}. The free energy of five-dimensional SCFTs has been computed and matched with results from the gravitational dual in \cite{Jafferis:2012iv}.

Another large class of five-dimensional SCFTs involves utilizing 5-brane webs within type IIB setups, \cite{Aharony:1997ju,Kol:1997fv,Aharony:1997bh}. These 5-brane webs offer a means to engineer infinite families of five-dimensional SCFTs, a framework that can be extended further by introducing 7-branes, \cite{DeWolfe:1999hj}. Corresponding to this diverse array of field theories, one anticipates the existence of extensive classes of $AdS_6$ solutions and their corresponding AdS$_6$/CFT$_5$ dualities within type IIB supergravity. Such classes of $AdS_6$ solutions have been systematically constructed in \cite{DHoker:2016ysh} and references thereof. The supergravity solutions exhibit geometries characterized by $AdS_6\times{S}^2$ warped over a Riemann surface, $\Sigma_\mathfrak{g}$. Notably, consistent truncations to $F(4)$ gauged supergravity have been established for such solutions, \cite{Hong:2018amk,Malek:2018zcz}, and these have been also extended to include additional vector multiplets, \cite{Malek:2019ucd}. The existence of these truncations implies that the AdS/CFT correspondence predicts a universal relationship between the five-sphere partition function and the topologically twisted index, akin to the findings for five-dimensional Seiberg theories, to similarly hold true for five-dimensional SCFTs endowed with holographic duals in type IIB supergravity, see \cite{Fluder:2019szh}.

As mentioned above, the theories of class $\mathcal{F}$ are three-dimensional SCFTs obtained through the twisted compactification of five-dimensional theories on a Riemann surface. Previously, the $AdS_4\times\Sigma_\mathfrak{g}$ solutions of $F(4)$ gauged supergravity, dual to these class $\mathcal{F}$ theories have been investigated in \cite{Nunez:2001pt, Naka:2002jz, Karndumri:2015eta, Hosseini:2018usu} where $\Sigma_\mathfrak{g}$ represents a Riemann surface of genus $\mathfrak{g}$. The exploration of $AdS_4$ solutions in massive type IIA supergravity which are dual to class $\mathcal{F}$ theories, was initiated in \cite{Bah:2018lyv}.

In this study, we focus on examining black hole solutions characterized by an orbifold horizon, believed to asymptote to the $AdS_4$ vacuum, serving as the dual description for three-dimensional class $\mathcal{F}$ theories on a circle times a spindle, $S^1\times\Sigma$.

In recent years, there has been extensive exploration into novel classes of twisted compactification on orbifolds. These orbifolds exhibit diverse structures, including spindle topologies resembling spheres with two orbifold singularities, as well as disk configurations featuring a single orbifold singularity. A wide array of studies, \cite{Ferrero:2020laf, Ferrero:2020twa, Hosseini:2021fge, Boido:2021szx, Ferrero:2021wvk, Cassani:2021dwa, Ferrero:2021ovq, Couzens:2021rlk, Faedo:2021nub, Ferrero:2021etw, Giri:2021xta, Couzens:2021cpk, Cheung:2022ilc, Suh:2022olh, Arav:2022lzo, Couzens:2022yiv, Couzens:2022aki, Couzens:2022lvg, Faedo:2022rqx, Suh:2022pkg, Suh:2023xse, Amariti:2023mpg, Kim:2023ncn, Hristov:2023rel, Amariti:2023gcx, Couzens:2023kyf, Faedo:2024upq, Ferrero:2024vmz, Boisvert:2024jrl}, along with the examination of defect solutions, \cite{Gutperle:2022pgw, Gutperle:2023yrd, Capuozzo:2023fll}, have contributed to our understanding of these intricate constructions.

To construct black hole solutions asymptotic to the $AdS_4$ vacuum, which serve as dual descriptions for three-dimensional class $\mathcal{F}$ theories, we delve into the consistent truncation of matter coupled $F(4)$ gauged supergravity, \cite{Andrianopoli:2001rs}, on a Riemann surface, \cite{Hosseini:2020wag}. This truncation yields $\mathcal{N}=2$ gauged supergravity coupled to two vector multiplets and a universal hypermultiplet in four dimensions. Within this framework, we find $AdS_2\times\Sigma$ solutions which represent the horizon geometry of presumed black hole solutions with $\Sigma$ denoting a spindle. To construct spindle solutions featuring a non-trivial hypermultiplet, we employ the methodology introduced in \cite{Arav:2022lzo} and further extended in subsequent works, \cite{Suh:2022pkg, Suh:2023xse, Amariti:2023mpg, Hristov:2023rel, Amariti:2023gcx}.

Upon uplift to six dimensions, our solutions become $AdS_2\times\Sigma\times\Sigma_\mathfrak{g}$ where $\Sigma$ represents a spindle and $\Sigma_\mathfrak{g}$ is a Riemann surface of genus $\mathfrak{g}$. After an additional uplift to massive type IIA supergravity, these solutions manifest as $AdS_2\times\Sigma\times\Sigma_\mathfrak{g}\times\widetilde{HS}^4$ configurations where $\widetilde{HS}^4$ denotes a squashed four-hemisphere. A similar uplift can be also considered in type IIB supergravity as explained above. Consequently, the $AdS_2\times\Sigma$ solutions we present here can be regarded as generalizations of previously known solutions, see below. 

It is important to note that our $AdS_2\times\Sigma$ solutions bear strong resemblance to the $AdS_3\times\Sigma$ solutions asymptotic to the $AdS_5$ vacuum studied in \cite{Amariti:2023gcx}. These $AdS_3\times\Sigma$ solutions were derived in $\mathcal{N}=2$ gauged supergravity coupled to vector multiplets and a hypermultiplet in five dimensions, via the consistent truncation of seven-dimensional gauged supergravity on a Riemann surface, \cite{Szepietowski:2012tb, MatthewCheung:2019ehr, Cassani:2019vcl, Faedo:2019cvr, Cassani:2020cod}, akin to the truncation in \cite{Hosseini:2020wag}.

Moreover, we perform calculations of the Bekenstein-Hawking entropy for the presumed black hole solutions. By leveraging gravitational blocks, \cite{Hosseini:2019iad, Hosseini:2021fge, Faedo:2021nub, Suh:2022olh, Faedo:2022rqx, Hristov:2023rel, Faedo:2024upq}, we compute the entropy and the values of scalar fields at the orbifold points, thus achieving precise agreement with the numeric results from the solutions.

\subsubsection*{Precise comparison with previous solutions}
We can be more quantitative in explaining the relation between our present findings and the already existing solutions in the literature. In order to do this, consider the simple fact that the twisted compactification on a Riemann surface from six to four dimensions features a pair of fluxes, $s^1$ and $s^2$, that obey the twist condition,
\be
	s^1 + s^2 = - \frac{\tilde{\kappa}}{3 m}\ ,
\ee
where $\tilde{\kappa}$ is the curvature of the Riemann surface we compactify on and $m$ is related to the cosmological constant in six dimensions and the Romans mass in ten dimensions. The fluxes, $s^1$ and $s^2$, in turn feature in the definition of the resulting four-dimensional model and it turns out that the limit of $s^1 = 0$ (equivalent due to symmetry to $s^2 = 0$) simplifies the model (it becomes equivalent to the so called $T^3$ model as explained in \cite{Hosseini:2020wag}) and allows for the analytic construction of spindle black holes, as done in \cite{Faedo:2021nub,Giri:2021xta,Suh:2022olh,Couzens:2022lvg,Faedo:2022rqx}. A generic value of $s^1$ instead leads to a more complicated model and is precisely the generalization we explore in the present work. Naturally all our results agree with the preexisting answers in the limit of vanishing $s^1$.

\subsubsection*{Outline of the paper}
In section \ref{sec:2}, we review $\mathcal{N}=2$ gauged supergravity coupled to vector multiplets and a hypermultiplet in four dimensions from the consistent truncation of $F(4)$ gauged supergravity on a Riemann surface. In section \ref{sec:3}, we derive and analyse the BPS equations for $AdS_2\times\Sigma$ solutions and calculate the Bekenstein-Hawking entropy. In section \ref{sec:4}, we solve the BPS equations and obtain numerical solutions of $AdS_2\times\Sigma$. In section \ref{sec:gluing} we switch gear and present the complementary description of our solutions in terms of gravitational blocks, drawing many relations with other recent approaches in the literature. In section \ref{sec:5} we conclude with outlook. Details of the consistent truncation and derivation of BPS equations are relegated in appendix \ref{appA} and \ref{appB}. In addition, we have included a complementary Mathematica notebook with the present submission, containing details on the numerical solutions and gravitational block matching and allowing one to change explicitly the various solution parameters and magnetic fluxes.

\section{The supergravity model}
\label{sec:2}

The bosonic Lagrangian was derived in  \cite{Hosseini:2020wag} by dimensional reduction on a Riemann surface of constant curvature $\tilde \kappa$. For the purposes of finding a static black hole spindle solution with no electric charges we directly perform a truncation to the axionless sector of the theory in appendix \ref{appA}. We are thus left with the metric, three $U(1)$ gauge fields $A_\mu^I$, $I=0,1,2$, two real vector multiplet scalar fields, $\chi_{1,2}$, and two real hypermultiplet scalar fields, $\phi$ and $\sigma$. The Lagrangian is given by
\begin{align} \label{lagmain}
e^{-1}\mathcal{L}\,&=\,\frac{1}{2}R-\partial_\mu\chi_1\partial^\mu\chi_1-\frac{1}{2}\partial_\mu\chi_2\partial^\mu\chi_2-\partial_\mu\phi\partial^\mu\phi-\frac{1}{4}e^{4\phi}D_\mu\sigma{}D^\mu\sigma-\mathcal{V} \notag \\
&-\frac{1}{8}\Big[2e^{2\chi_1}F^0_{\mu\nu}F^{0\mu\nu}+e^{-2\chi_1-2\chi_2}F^1_{\mu\nu}F^{1\mu\nu}+e^{-2\chi_1+2\chi_2}F^2_{\mu\nu}F^{2\mu\nu}\Big]\,,
\end{align}
where we define
\begin{equation}
D\sigma\,=\,d\sigma-4mA^0+s_2A^1+s_1A^2\,.
\end{equation}
The scalar potenatial,
\begin{equation}
\mathcal{V}\,=\,\left(\frac{\partial\,W}{\partial\,\chi_1}\right)^2+2\left(\frac{\partial\,W}{\partial\,\chi_2}\right)^2+\left(\frac{\partial\,W}{\partial\phi}\right)^2-3W^2\,,
\end{equation}
is given in terms of the superpotential,
\begin{equation} \label{superpot}
W\,=\,3me^{\chi_1}\cosh\chi_2+\frac{1}{4}e^{2\phi}\left(4me^{-\chi_1}-s_1e^{\chi_1-\chi_2}-s_2e^{\chi_1+\chi_2}\right)\,,
\end{equation}
where $s_1$ and $s_2$ are free parameters from the magnetic charges in $F(4)$ gauged supergravity,
\begin{equation}
s_1+s_2\,=\,-\frac{\tilde{\kappa}}{3m}\,,
\end{equation}
and $m$ is the gauge coupling and $\tilde{\kappa}=\pm1$ is the curvature of Riemann surface on which $F(4)$ gauged supergravity is truncated.

The supersymmetry variations of gravitino, gaugino and hyperino are 
\begin{align}
\Big[2\nabla_\mu-iB_\mu-W\gamma_\mu+4iH_{\mu\nu}\gamma^\nu\Big]\epsilon\,=&\,0\,, \notag \\
\Big[\partial_\mu\chi_1\gamma^\mu+\partial_{\chi_1}W+i\partial_{\chi_1}H_{\mu\nu}\gamma^{\mu\nu}\Big]\epsilon\,=&\,0\,, \notag \\
\Big[\frac{1}{2}\partial_\mu\chi_2\gamma^\mu+\partial_{\chi_2}W+i\partial_{\chi_2}H_{\mu\nu}\gamma^{\mu\nu}\Big]\epsilon\,=&\,0\,, \notag \\
\left[\partial_\mu\phi\gamma^\mu+\partial_\phi{W}+\frac{i}{2}\partial_\phi{B}_\mu\gamma^\mu\right]\epsilon\,=&\,0\,,
\end{align}
where we introduce
\begin{align} \label{hhbbdef}
H_{\mu\nu}\,=&\,-\frac{1}{2}\left(2e^{\chi_1}F^{-0}_{\mu\nu}+e^{-\chi_1-\chi_2}F^{-1}_{\mu\nu}+e^{-\chi_1+\chi_2}F^{-2}_{\mu\nu}\right)\,, \notag \\
B_\mu\,=&\,-3m\left(A^1_\mu+A^2_\mu\right)+\frac{1}{2}e^{2\phi}D_\mu\sigma\,.
\end{align}
We introduce a real parameter, $z$, for $s_1$ and $s_2$,
\begin{equation}
s_1\,=\,-\frac{\tilde{\kappa}}{6m}\left(1+\frac{z}{\tilde{\kappa}}\right)\,, \qquad s_2\,=\,-\frac{\tilde{\kappa}}{6m}\left(1-\frac{z}{\tilde{\kappa}}\right)\,.
\end{equation}
The scalar potential has a vacuum at, \cite{Hosseini:2020wag},
\begin{align} \label{radii}
\chi_{1*}\,=&\,\frac{1}{4}\log\left(\frac{288m^4}{2z^2+\tilde{\kappa}\left(\tilde{\kappa}-\sqrt{8z^2+\tilde{\kappa}^2}\right)}\right)\,, \notag \\
\chi_{2*}\,=&\,\frac{1}{2}\log\left(\frac{z+\tilde{\kappa}}{3z-\sqrt{8z^2+\tilde{\kappa}^2}}\right)\,, \notag \\
\phi_*\,=&\,\frac{1}{2}\log\left(\frac{72m^2}{-3\tilde{\kappa}+\sqrt{8z^2+\tilde{\kappa}^2}}\right)\,.
\end{align}
We assumed $s_1$ to be non-vanishing or, equivalently, $\tilde{\kappa}z\ne-1$. The scalar field, $\sigma$, is a flat direction and, by the Higgs mechanism, gets eaten by the massive vector field, $A^m$. At the vacuum, the value of scalar potential is
\begin{equation} \label{potads4}
\mathcal{V}_*\,=\,-\frac{3}{L_{AdS_4}^2}\,=\,-2\left(6m\right)^4\frac{\sqrt{4z^2+2\tilde{\kappa}\left(\tilde{\kappa}-\sqrt{\tilde{\kappa}^2+8z^2}\right)}}{\left(3\tilde{\kappa}-\sqrt{\tilde{\kappa}^2+8z^2}\right)^2}\,.
\end{equation}
For $\tilde{\kappa}=+1$, we require $z^2>1$ and any $z$ is allowed for $\tilde{\kappa}=-1$, \cite{Bah:2018lyv}. The R-symmetry, massive and flavor vector fields are given by, respectively,
\begin{align} \label{vectorsrmbb}
A^R\,=&\,3m\left(A^1+A^2\right)\,, \notag \\
A^m\,=&\,-4mA^0+s_2A^1+s_1A^2\,, \notag \\
A^F\,=&\,3m\left(e^{-\chi_{1*}-\chi_{2*}}A^1-e^{-\chi_{1*}+\chi_{2*}}A^2\right)\,,
\end{align}
where $\chi_{i*}$ are the values of the scalar fields at the vacuum in \eqref{radii}.

By the AdS/CFT correspondence, the free energy of pure $AdS_4$ with an asymptotic boundary of $S^3$ is
\begin{equation} \label{flads4}
\mathcal{F}_{S^3}\,=\,\frac{\pi{L}_{AdS_4}^2}{2G_N^{(4)}}\,,
\end{equation}
where $G_N^{(4)}$ is the four-dimensional Newton's gravitational constant. The free energy of class $\mathcal{F}$ theories (setting $m = 1/2$) is given by, \cite{Hosseini:2020wag},
\begin{equation} \label{fabjm}
\mathcal{F}_{S^3}^{\text{class} \mathcal{F}}\,=\,-\frac{1}{9}|1-\mathfrak{g}|\frac{\left(3\tilde{\kappa}-\sqrt{\tilde{\kappa}^2+8z^2}\right)^2}{\sqrt{4z^2+2\tilde{\kappa}\left(\tilde{\kappa}-\sqrt{\tilde{\kappa}^2+8z^2}\right)}}\mathcal{F}_{S^5}\,,
\end{equation}
where the four-dimensional Newton's constant relates to the five-sphere free energy of the dual field theory via the following two identities,
\be
	\frac1{G_N^{(6)}} = - \frac3{\pi^2}\, \cF_{S^5}\ , \qquad \frac1{G_N^{(4)}} = \frac{4\pi |1-\frak{g}|}{G_N^{(6)}} = - \frac{12}{\pi}\, |1-\frak{g}|\, \cF_{S^5}\ .
\ee
The supergravity theory derived in \cite{Hosseini:2020wag} is believed to be embeddable in either type IIA or type IIB supergravity in ten dimensions allowing the description of several different field theory duals as outlined in the introduction. The additional $U(1)$ flavor symmetry is however most directly identifiable in the IIA description, dual to the so-called Seiberg theories \cite{Seiberg:1996bd,Brandhuber:1999np}, see \cite{Hosseini:2018usu} for direct evidence of this claim. In this case we consider the free energy of 5d $USp(2N)$ theories with $N_f$ flavors,
\begin{equation}
\mathcal{F}_{S^5}^{\text{Seiberg}}\,=\,-\frac{9\sqrt{2}\pi}{5}\frac{N^{5/2}}{\sqrt{8-N_f}}\,.
\end{equation}

\section{$AdS_2$ ansatz}
\label{sec:3}

We consider the metric and the gauge fields,
\begin{align}
ds^2\,=&\,e^{2V}ds_{AdS_2}^2+f^2dy^2+h^2dz^2\,, \notag \\
A^I\,=&\,a^Idz\,,
\end{align}
where $ds^2_{AdS_2}$ is a unit radius metric on $AdS_2$ and $V$, $f$, $h$, and $a^I$, $I=0,1,2$, as well as the scalar fields $\phi, \chi_i$, $i=1,2$, are functions of $y$-coordinate only. In order to avoid partial differential equations from the equations of motion for the gauge fields, the scalar field, $\sigma$, is given by $\sigma=\bar{\sigma}z$ where $\bar{\sigma}$ is constant. Hence, we find
\begin{equation}
B_\mu{d}x^\mu\,\equiv\,B_zdz\,,
\end{equation}
where $B_z$ is again a function of the $y$-coordinate only.

We employ an orthonormal frame,
\begin{equation}
e^a\,=\,e^V\bar{e}^a\,, \qquad e^2\,=\,fdy\,, \qquad e^3\,=\,hdz\,,
\end{equation}
where $\bar{e}^a$ is an orthonormal frame on $ds_{AdS_2}^2$. In the frame coordinates, the field strengths are given by
\begin{equation}
F_{23}^I\,=\,f^{-1}h^{-1}\left(a^I\right)'\,.
\end{equation}

From the Maxwell equations we find two independent integrals of motion,
\begin{align} \label{er123}
-2ms_2e^{2V}\left(e^{2\chi_1}\frac{1}{2m}F_{23}^0+e^{-2\chi_1-2\chi_2}\frac{1}{s_2}F_{23}^1\right)\,=\,\mathcal{E}_{R_1}\,, \notag \\
-2ms_1e^{2V}\left(e^{2\chi_1}\frac{1}{2m}F_{23}^0+e^{-2\chi_1+2\chi_2}\frac{1}{s_1}F_{23}^2\right)\,=\,\mathcal{E}_{R_2}\,,
\end{align}
with
\begin{align}
\left(2e^{2V+2\chi_1}F^0_{23}\right)'\,=&\,4me^{2V}fh^{-1}e^{4\phi}D_z\sigma\,, \notag \\
\left(e^{2V-2\chi_1-2\chi_2}F^1_{23}\right)'\,=&\,-s_2e^{2V}fh^{-1}e^{4\phi}D_z\sigma\,, \notag \\
\left(e^{2V-2\chi_1+2\chi_2}F^2_{23}\right)'\,=&\,-s_1e^{2V}fh^{-1}e^{4\phi}D_z\sigma\,,
\end{align}
where $\mathcal{E}_{R_i}$ are constant.

\subsection{BPS equations}
\label{sec:3.1}

We choose the gamma matrix to be
\begin{equation}
\gamma^m\,=\,\Gamma^m\otimes\sigma^3\,, \qquad \gamma^2\,=\,I_2\otimes\sigma^1\,, \qquad \gamma^3\,=\,I_2\otimes\sigma^2\,,
\end{equation}
where $\Gamma^m$ are two-dimensional gamma matrices and $\sigma^i$ the Pauli matrices. The spinors are
\begin{equation}
\epsilon\,=\,\psi\otimes\chi\,.
\end{equation}
The two-dimensional spinor satisfies
\begin{equation}
D_m\psi\,=\,\frac{1}{2}\kappa\Gamma_m\psi\,,
\end{equation}
where $\kappa=\pm1$ fixes the chirality.

For $\sin\xi\ne0$ where $\xi$ is an angular parameter introduced in \eqref{intxi}, we present the complete BPS equations,
\begin{align} \label{fullbps}
f^{-1}\xi'\,=&\,2W\cos\xi+\kappa{e}^{-V}\,, \notag \\
f^{-1}V'\,=&\,W\sin\xi\,, \notag \\
f^{-1}\chi_1'\,=&\,-\partial_{\chi_1}W\sin\xi\,, \notag \\
f^{-1}\chi_2'\,=&\,-2\partial_{\chi_2}W\sin\xi\,, \notag \\
f^{-1}\phi'\,=&\,-\frac{\partial_\phi{W}}{\sin\xi}\,, \notag \\
f^{-1}\frac{h'}{h}\,=&\,\frac{1}{\sin\xi}\Big(\kappa{e}^{-V}\cos\xi+W\left(1+\cos^2\xi\right)\Big)\,,
\end{align}
with two constraints,
\begin{align} \label{bpsconstraints}
\left(s-B_z\right)\sin\xi\,=&\,-2Wh\cos\xi-\kappa{h}e^{-V}\,, \notag \\
\partial_\phi{W}\cos\xi\,=&\,\frac{1}{2}\partial_\phi{B}_z\sin\xi{h}^{-1}\,.
\end{align}
See appendix \ref{appB} for more details on the BPS equations. The field strengths of gauge fields are
\begin{align} \label{bpsfs}
\partial_{\chi_1}H_{23}\,=&\,-\frac{1}{4}\partial_{\chi_1}W\cos\xi\,, \notag \\
\partial_{\chi_2}H_{23}\,=&\,-\frac{1}{4}\partial_{\chi_2}W\cos\xi\,, \notag \\
H_{23}\,=&\,-\frac{1}{4}W\cos\xi-\frac{1}{4}\kappa{e}^{-V}\,.
\end{align}
The BPS equations are consistent with the equations of motion from the Lagrangian in \eqref{lagmain}.

\subsection{Integrals of motion}

We find an integral of the BPS equations,
\begin{equation} \label{hevks}
he^{-V}\,=\,k\sin\xi\,,
\end{equation}
where $k$ is a constant. Thus, at the poles of the spindle solutions at $h=0$, we find $\sin\xi=0$. From \eqref{fullbps} and \eqref{bpsconstraints} we obtain
\begin{equation} \label{xipxip}
\xi'\,=\,-k^{-1}\left(s-B_z\right)\left(e^{-V}f\right)\,,
\end{equation}
and the two constraints in \eqref{bpsconstraints} become
\begin{align} \label{constrainttwo}
\left(s-B_z\right)\,=&\,-k\Big[2We^V\cos\xi+\kappa\Big]\,, \notag \\
\partial_\phi{W}\cos\xi\,=&\,\frac{1}{2}k^{-1}e^{-V}\partial_\phi{B}_z\,.
\end{align}

From the field strengths in \eqref{bpsfs}, we find expressions of the integrals of motion to be
\begin{align}
\mathcal{E}_{R_1}\,=&\,e^V\left[12m^2e^V\cos\xi+\kappa\left(2me^{-\chi_1-\chi_2}+s_2e^{\chi_1}\right)\right]\,, \notag \\
\mathcal{E}_{R_2}\,=&\,e^V\left[12m^2e^V\cos\xi+\kappa\left(2me^{-\chi_1+\chi_2}+s_1e^{\chi_1}\right)\right]\,.
\end{align}

\subsection{Boundary conditions for spindle solutions} \label{sec33}

We employ the conformal gauge,
\begin{equation}
f\,=\,-e^V\,,
\end{equation}
in order to have the metric in the form of
\begin{equation}
ds^2\,=\,e^{2V}\left[ds_{AdS_2}^2+ds_{\Sigma}^2\right]\,,
\end{equation}
where the metric of the spindle is
\begin{equation}
ds_{\Sigma}^2\,=\,dy^2+k^2\sin^2\xi{d}z^2\,.
\end{equation}
The spindle solutions have two poles at $y=y_{N,S}$ with deficit angles of $2\pi\left(1-\frac{1}{n_{N,S}}\right)$. The period of the azimuthal angle, $z$, is set to be
\begin{equation}
\Delta{z}\,=\,2\pi\,.
\end{equation}

\subsubsection{Analysis of the BPS equations}

We analyze the BPS equations for the spindle solutions. At the poles, $y\,=\,y_{N,S}$, as $k\sin\xi\rightarrow0$, we find $\cos\xi\rightarrow\pm1$ if $k\ne0$. Thus, we obtain $\cos\xi_{N,S}\,=\,(-1)^{t_{N,S}}$ with $t_{N,S}\in\{0,1\}$. We choose $y_N<y_S$ and $y\in[y_N,y_S]$. We assume the deficit angles at the poles to be $2\pi\left(1-\frac{1}{n_{N,S}}\right)$ with $n_{N,S}\ge1$. Then we require the metric to have $|\left(k\sin\xi\right)'|_{N,S}\,=\,\frac{1}{n_{N,S}}$. By the symmetry of the BPS equations in \eqref{hsymm} and \eqref{hevks}, we further choose
\begin{equation}
h\le0\,, \qquad \Leftrightarrow \qquad k\sin\xi\le0\,.
\end{equation}
Then we find $\left(k\sin\xi\right)'|_N<0$ and $\left(k\sin\xi\right)'|_S>0$. Thus, we impose 
\begin{equation}
\left(k\sin\xi\right)'|_{N,S}\,=\,-\frac{(-1)^{l_{N,S}}}{n_{N,S}}\,, \qquad l_N\,=\,0\,,l_S\,=\,1\,.
\end{equation}

For the spindle solutions, we have two distinct classes, the twist and the anti-twist classes, \cite{Ferrero:2021etw}. The spinors are of the same chirality at the poles for twist solutions and opposite chiralities for anti-twist solutions,
\begin{align} \label{cosxi1}
\cos\xi|_{N,S}\,=\,(-1)^{t_{N,S}}; \qquad &\text{Twist:} \,\, \qquad \quad \left(t_N,t_S\right)\,=\,\left(1,1\right) \quad \text{or} \quad \left(0,0\right)\,, \notag \\
&\text{Anti-Twist:} \quad \left(t_N,t_S\right)\,=\,\left(1,0\right) \quad \text{or} \quad \left(0,1\right)\,.
\end{align}

As we have $\left(k\sin\xi\right)'\,=\,+\cos\xi\left(s-B_z\right)$ from the BPS equation in \eqref{xipxip}, we obtain
\begin{equation}
\left(s-B_z\right)|_{N,S}\,=\,\frac{1}{n_{N,S}}(-1)^{l_{N,S}+t_{N,S}+1}\,.
\end{equation}
We consider the flux quantization for R-symmetry flux. From \eqref{hhbbdef} we find $-F^R\,=\,dB-d\left(\frac{1}{2}e^{2\phi}D\sigma\right)$. At the poles, as $\phi=0$ unless $D\sigma=0$, the second term on the right hand side of $F^R$ does not contribute to the flux quantization. Then, the R-symmetry flux quantized is given by
\begin{equation} \label{rsymmq}
\frac{1}{2\pi}\int_{{\Sigma}}F^R\,\equiv\,\frac{1}{2\pi}\int_{\Sigma}\left(-dB\right)\,=\,\frac{n_N(-1)^{t_S}+n_S(-1)^{t_N}}{n_Nn_S}\,.
\end{equation}

We have $\partial_zB=e^{2\phi}D_z\sigma$. Again, as $\phi=0$ unless $D\sigma=0$ at the poles, we obtain $\partial_\phi{B}_z=0$ at the poles. We also find $\partial_\phi{W}=0$ at the poles from the constraint in \eqref{constrainttwo}. Thus, we obtain
\begin{equation} \label{dbdw}
\partial_\phi{B}_z|_{N,S}\,=\,\partial_\phi{W}|_{N,S}\,=\,0\,.
\end{equation}

We further make assumption that the hypermultiplet scalar fields, $(\phi,\sigma)$, are non-vanishing at the poles and we obtain
\begin{equation} \label{chitopsi}
\phi|_N\,,\phi|_S\,\ne\,0\,, \qquad \Rightarrow \qquad D_z\sigma|_N\,=\,D_z\sigma|_S\,=\,0\,.
\end{equation}
Hence, the flux charging $\sigma$ should vanish,
\begin{equation} \label{msymmq}
\frac{1}{2\pi}\int_{\Sigma}\left(-4mF^0+s_2F^1+s_1F^2\right)\,=\,\left(D_z\sigma\right)|_{y_N}^{y_S}\,=\,0\,.
\end{equation}

From \eqref{chitopsi} and the second equation in \eqref{dbdw} we obtain
\begin{equation} \label{dw=0}
\left.\left(4me^{-\chi_1}-s_2e^{\chi_1+\chi_2}-s_1e^{\chi_1-\chi_2}\right)\right|_{N,S}\,=\,0\,, \quad \Rightarrow \quad W|_{N,S}\,=\,3me^{\chi_1}\cosh\chi_2|_{N,S}\,.
\end{equation}
We define a quantity,
\begin{equation} \label{defm1m2}
M_{(1)}\,\equiv\,-3me^Ve^{\chi_1}\cosh\chi_2\,,
\end{equation}
where $M_{(1)}<0$. By the first equation in \eqref{constrainttwo} we eliminate $V$ and by \eqref{cosxi1} eliminate $\cos\xi$. Then we find the integrals of motion to be
\begin{align}
\mathcal{E}_{R_1}\,=\,-\frac{\kappa}{3m}M_{(1)}\frac{2me^{-2\chi_1-\chi_2}+s_2}{\cosh\chi_2}+\frac{4M^2_{(1)}}{3}\cos\xi\frac{e^{-2\chi_2}}{\cosh^2\chi_2}\,, \notag \\
\mathcal{E}_{R_2}\,=\,-\frac{\kappa}{3m}M_{(1)}\frac{2me^{-2\chi_1+\chi_2}+s_1}{\cosh\chi_2}+\frac{4M^2_{(1)}}{3}\cos\xi\frac{e^{-2\chi_2}}{\cosh^2\chi_2}\,,
\end{align}
where we have
\begin{equation}
M_{(1)}\,=\,\frac{1}{2}(-1)^{t_{N,S}}\kappa-\frac{1}{2kn_{N,S}}(-1)^{l_{N,S}}\,.
\end{equation}
Finally, one of the scalar fields, $\chi_2$, can be eliminated by employing the condition on the left hand side of \eqref{dw=0}. Then as the integrals of motion are constant and have identical values at the poles, we can solve for $\left(\chi_{1N},\chi_{1S}\right)$ from
\begin{align} \label{assoeq}
\mathcal{E}_{R_1}\left(\chi_{1N}\right)\,=&\,\mathcal{E}_{R_1}\left(\chi_{1S}\right)\,, \notag \\
\mathcal{E}_{R_2}\left(\chi_{1N}\right)\,=&\,\mathcal{E}_{R_2}\left(\chi_{1S}\right)\,.
\end{align}
This fixes all the values of the functions at the poles beside the scalar field, $\phi$, from the hypermultiplet.

\subsubsection{Fluxes}

In appendix \ref{appB}, we have expressed the field strengths in terms of the scalar fields, metric functions, the angle, $\xi$, and constant, $k$,
\begin{equation}
F^I_{yz}\,=\,\left(a^I\right)'\,=\,\left(\mathcal{I}^{(I)}\right)'\,,
\end{equation}
where we define
\begin{align} \label{idef}
\mathcal{I}^{(0)}\,\equiv&\,\frac{1}{\sqrt{2}}ke^V\cos\xi{e}^{-\chi_1}\,, \notag \\
\mathcal{I}^{(1)}\,\equiv&\,\frac{1}{\sqrt{2}}ke^V\cos\xi{e}^{\chi_1+\chi_2}\,, \notag \\
\mathcal{I}^{(2)}\,\equiv&\,\frac{1}{\sqrt{2}}ke^V\cos\xi{e}^{\chi_1-\chi_2}\,.
\end{align}
By the data at the poles, the fluxes are solely determined ,
\begin{equation}
\frac{p_I}{n_Nn_S}\,\equiv\,\frac{1}{2\pi}\int_{\Sigma}F^I\,=\,\mathcal{I}_I|_N^S\,.
\end{equation}

From \eqref{vectorsrmbb} we define R-symmetry, massive and flavor vector fluxes, respectively,
\begin{align}
\mathcal{I}_R|_{N,S}\,\equiv&\,\left.3m\left(\mathcal{I}^{(1)}+\mathcal{I}^{(2)}\right)\right|_{N,S}\,, \notag \\
\mathcal{I}_m|_{N,S}\,\equiv&\,-4m\mathcal{I}^{(0)}+s_2\mathcal{I}^{(1)}+s_1\mathcal{I}^{(2)}|_{N,S}\,, \notag \\
\mathcal{I}_F|_{N,S}\,\equiv&\,\left.3m\left(e^{-\chi_{1*}-\chi_{2*}}\mathcal{I}^{(1)}-e^{-\chi_{1*}+\chi_{2*}}\mathcal{I}^{(2)}\right)\right|_{N,S}\,,
\end{align}
where $\chi_{i*}$ are the values of the scalar fields at the vacuum in \eqref{radii}. From \eqref{idef} and \eqref{dw=0}, we find
\begin{align}
\mathcal{I}_R|_{N,S}\,=&\,3m\left(\mathcal{I}^{(1)}+\mathcal{I}^{(2)}\right)|_{N,S} \notag \\
=&\,2k\cos\xi\,3me^Ve^{\chi_1}\cosh\chi_2|_{N,S} \notag \\
=&\,2kM_{(1)}|_{N,S}(-1)^{t_{N,S}}\,, \\
\notag \\
\mathcal{I}_m|_{N,S}\,=&\,-4m\mathcal{I}^{(0)}+s_2\mathcal{I}^{(1)}+s_1\mathcal{I}^{(2)}|_{N,S} \notag \\
=&\,ke^V\cos\xi\left(-4me^{-\chi_1}+s_2e^{\chi_1+\chi_2}+s_1e^{\chi_1-\chi_2}\right)|_{N,S}\,=\,0\,,
\end{align}
where we employed \eqref{dw=0}. Then we recover the R-symmetry flux quantization, \eqref{rsymmq}, and the vanishing of the flux of massive vector field, \eqref{msymmq}, respectively,
\begin{align}
\mathcal{I}_R|_N^S\,=&\,\frac{n_N(-1)^{t_S}+n_S(-1)^{t_N}}{n_Nn_S}\,, \notag \\
\mathcal{I}_m|_N^S\,=&\,0\,.
\end{align}
The flux of flavor vector field is
\begin{align} \label{fsymmq1}
\frac{p_F}{n_Nn_S}\,\equiv&\,\mathcal{I}_F|_N^S\,=\,3m\left(e^{-\chi_{1*}-\chi_{2*}}\mathcal{I}^{(1)}-e^{-\chi_{1*}+\chi_{2*}}\mathcal{I}^{(2)}\right)|_N^S \notag \\
=&\,\left.kM_{(1)}(-1)^{t_{N,S}}\frac{e^{-\chi_{1*}-\chi_{2*}}e^{\chi_{1}+\chi_{2}}-^{-\chi_{1*}+\chi_{2*}}e^{\chi_{1}-\chi_{2}}}{e^{\chi_1}\cosh\chi_2}\right|_N^S\,,
\end{align}
where $p_F$ is an integer. The expression of $k$ is determined by this constraint.

{\bf Summary of the constraints to determine all the boundary conditions:} We summarize the constraints obtained to determine all the boundary conditions. By solving six associated equations, the left hand side of \eqref{dw=0}, \eqref{assoeq}, and \eqref{fsymmq1}, we can determine the values of the scalar fields, $\chi_1$ and $\chi_2$ at the north and south poles and also the constant, $k$, in terms of $n_{N,S}$, $t_{N,S}$, $p_F$. Then the values of the metric function, $V$, at the poles are determined from the definition of $M_{(1)}$ in \eqref{defm1m2}. This fixes all the boundary conditions except the scalar field, $\phi$, from the hypermultiplet which will be chosen when constructing the solutions explicitly. However, the constraint equations are quite complicated and it appears to be not easy to solve them.

Even though we are not able to solve for the boundary conditions in terms of $n_{N,S}$, $t_{N,S}$, and $p_F$ analytically, if we choose numerical values of $n_{N,S}$, $t_{N,S}$, and $p_F$, the constraints can be solved to determine all the boundary conditions. For instance, in the anti-twist class, for the choice of
\begin{align} \label{ninput}
z\,=&\,0\,, \qquad \tilde{\kappa}\,=\,-1\,, \notag \\
n_N\,=&\,3\,, \qquad n_S\,=\,2\,, \qquad p_F\,=\,0.5\,, \notag \\
m\,=&\,1/2\,, \qquad \kappa\,=\,-1\,,
\end{align}
we find the boundary conditions to be
\begin{align} \label{noutput}
e^{\chi_{1N}}\,\approx&\,1.25405\,, \qquad \,\,\,\,\,\, e^{\chi_{1S}}\,\approx\,1.18731\,, \notag\\
e^{\chi_{2N}}\,\approx&\,3.53213\,, \qquad \quad \,\,\, e^{\chi_{2S}}\,\approx\,4.00666\,, \notag\\
k\,\approx&\,0.42271\,.
\end{align}
In this way, without finding analytic expression of the Bekenstein-Hawking entropy, we can determine numerical value for each choice of $n_{N,S}$, $t_{N,S}$, and $p_F$. Furthermore, we will be able to construct the solutions explicitly numerically, see below and in the attached {\it Mathematica} file.

\subsubsection{The Bekenstein-Hawking entropy} \label{sec333}

We compute the Bekenstein-Hawking entropy of the presumed black hole solution which asymptotes to the $AdS_4$ vacuum, \eqref{radii}, dual to the theories of class $\mathcal{F}$.

The AdS/CFT dictionary, \eqref{flads4} and \eqref{fabjm}, gives the four-dimensional Newton's constant,
\begin{equation}
\frac{1}{2G_N^{(4)}}\,=\,\frac{2\sqrt{2}\left(6m\right)^4|1-\mathfrak{g}|}{15}\frac{N^{5/2}}{\sqrt{8-N_f}}\,.
\end{equation}
Then the two-dimensional Newton's constant is
\begin{equation}
\left(G_N^{(2)}\right)^{-1}\,=\,\left(G_N^{(4)}\right)^{-1}\Delta{z}\int_{y_N}^{y_S}|fh|dy\,.
\end{equation}
Employing the BPS equations, we find
\begin{equation}
fh\,=\,ke^Vf\sin\xi\,=\,+\frac{k}{\kappa}\left(e^{2V}\cos\xi\right)'\,.
\end{equation}
Thus the Bekenstein-hawking entropy is solely expressed by the data at the poles,
\begin{align} \label{bhent}
&S_{\text{BH}}\,=\,\frac{1}{4G_N^{(2)}}\,=\,\frac{2\sqrt{2}\pi\left(6m\right)^4|1-\mathfrak{g}|}{15}\frac{N^{5/2}}{\sqrt{8-N_f}}\left(-\frac{k}{\kappa}\right)\left[e^{2V}\cos\xi\right]_N^S \notag \\
&=\,-\frac{2\sqrt{2}\left(6m\right)^4|1-\mathfrak{g}|}{15}\frac{N^{5/2}}{\sqrt{8-N_f}}\frac{k}{\kappa}\left(\left.\frac{M_{(1)}^2(-1)^{t_S}}{9m^2e^{2\chi_1}\cosh^2\chi_2}\right|_S-\left.\frac{M_{(1)}^2(-1)^{t_N}}{9m^2e^{2\chi_1}\cosh^2\chi_2}\right|_N\right)\,.
\end{align}

As we can determine the numerical values of the boundary conditions for each choice of $n_{N,S}$, $t_{N,S}$, and $p_F$, we can find the numerical value of the Bekenstein-Hawking entropy as well. For instance, for the choice of \eqref{ninput}, the Bekenstein-Hawking entropy is given by $S_{\text{BH}}\,\approx\,0.0294064|1-\mathfrak{g}|\frac{N^{5/2}}{\sqrt{8-N_f}}$.

Furthermore, when there is no flavor charge, $p_F=0$, we perform a non-trivial check that the numerical value of the Bekenstein-Hawking entropy precisely matches the value obtained from the formula given in \eqref{minent} for the solutions from minimal gauged supergravity.

\section{Solving the BPS equations} \label{sec:4}

\subsection{Analytic solutions for minimal gauged supergravity} \label{sec41}

In minimal gauged supergravity associated with the $AdS_4$ vacuum in \eqref{radii} dual to the theories of class $\mathcal{F}$, utilizing the $AdS_2\times\Sigma$ solutions in \cite{Ferrero:2020twa}, we find solutions in the anti-twist class to the BPS equations in \eqref{fullbps}, \eqref{bpsconstraints} and \eqref{bpsfs}. The scalar fields take the value at the $AdS_4$ vacuum in \eqref{radii}. The metric and the gauge field are
\begin{align}
ds^2\,=&\,L_{AdS_4}^2\left[\frac{y^2}{4}ds_{AdS_2}^2+\frac{y^2}{q(y)}dy^2+\frac{q(y)}{4y^2}c_0^2dz^2\right]\,, \notag \\
e^{\chi_{1*}}A^0\,=&\,e^{-\chi_{1*}-\chi_{2*}}A^1\,=\,e^{-\chi_{1*}+\chi_{2*}}A^2\,=\,-\left[\frac{c_0\kappa}{2}\left(1-\frac{a}{y}\right)+s\right]dz\,,
\end{align}
and where $\chi_{i*}$ are the values of the scalar fields at the vacuum in \eqref{radii} and we have
\begin{equation}
\sin\xi\,=\,-\frac{\sqrt{q(y)}}{y^2}\,, \qquad \cos\xi\,=\,\kappa\frac{2y-a}{y^2}\,.
\end{equation}
Note that for the overall factor in the metric, we have $L_{AdS_4}^2$ for the $AdS_4$ vacuum from \eqref{potads4}. The quartic function is given by
\begin{equation}
q(y)\,=\,y^4-4y^2+4ay-a^2\,,
\end{equation}
and the constants are
\begin{align}
a\,=&\,\frac{n_S^2-n_N^2}{n_S^2+n_N^2}\,, \notag \\
c_0\,=&\,\frac{\sqrt{n_S^2+n_N^2}}{\sqrt{2}n_Sn_N}\,.
\end{align}
We set $n_S>n_N$. For the two middle roots of $q(y)$, $y\in[y_N,y_S]$, we find
\begin{equation}
y_N\,=\,-1+\sqrt{1+a}\,, \qquad y_S\,=\,1-\sqrt{1-a}\,.
\end{equation}
The Bekenstein-Hawking entropy is calculated to give
\begin{align} \label{minent}
S_{\text{BH}}\,=&\,\frac{\sqrt{2}\sqrt{n_S^2+n_N^2}-\left(n_S+n_N\right)}{n_Sn_N}\frac{\pi{L}_{AdS_4}^2}{4G_N^{(4)}} \notag \\
=&\,\frac{\sqrt{2}\sqrt{n_S^2+n_N^2}-\left(n_S+n_N\right)}{n_Sn_N}\frac{1}{2}\mathcal{F}_{S^3}^{\text{class}\mathcal{F}}\,,
\end{align}
where we employed \eqref{radii} and \eqref{flads4} and $\mathcal{F}_{S^3}^{\text{class}\mathcal{F}}$, is the free energy of the theories of class $\mathcal{F}$ in \eqref{fabjm}.

\subsection{Numerical solutions for $p_F\ne0$}

In section \ref{sec33}, although we were not able to find the analytic expressions of the boundary conditions, we were able to determine the numerical values of the boundary conditions for each choice of $n_{N,S}$, $t_{N,S}$, $p_F$. Employing these results for the boundary conditions, we can numerically construct $AdS_2\times\Sigma$ solutions in the anti-twist class by solving the BPS equations.~{\footnote{As we do not know the analytic expressions of the boundary conditions, we could not exclude the existence of solutions in the twist class. However, we were not able to find any boundary conditions for numerical solutions in the twist class. In fact, we found solutions with the Bekenstein-Hawking entropy matching the result of gravitational block calculations. However, the scalar fields of the solutions were not real.}}

In order to solve the BPS equations numerically, we start the integration at $y=y_N$ and we choose $y_N=0$. At the poles we have $\sin\xi=0$. We scan over the initial value of $\phi$ at $y=y_N$ in search of a solution for which we have $\sin\xi=0$ in a finite range, $i.e.$, at $y=y_S$. If we find compact spindle solution, our boundary conditions guarantee the fluxes to be properly quantized.

\begin{figure}[t]
\begin{center}
\includegraphics[width=2.8in]{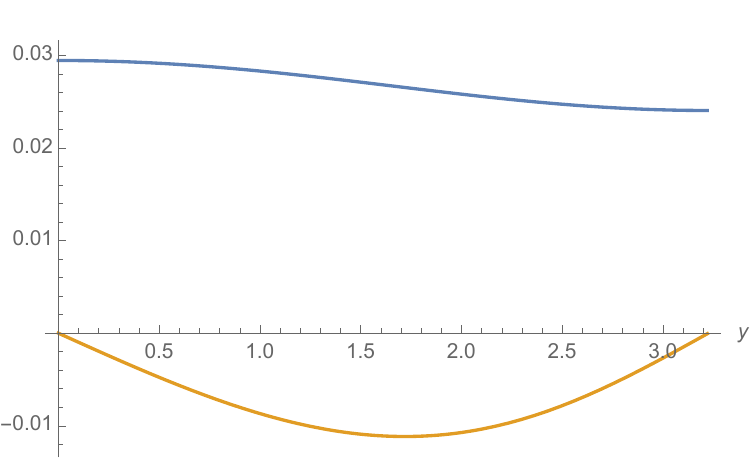} \qquad \includegraphics[width=2.8in]{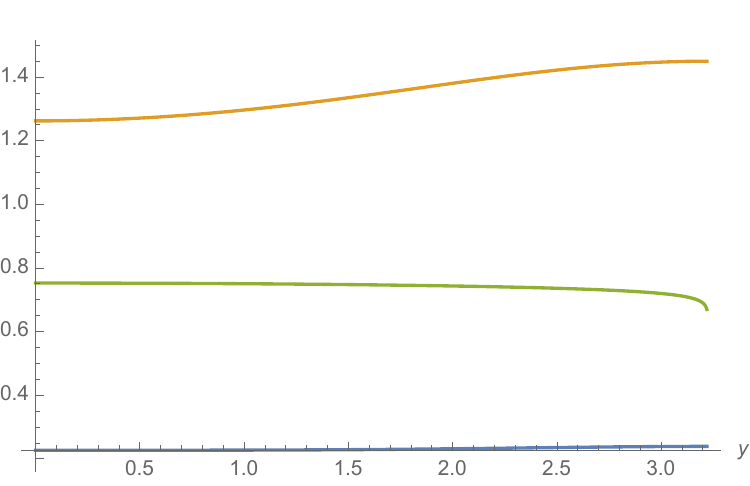}
\caption{{\it A representative $AdS_2\times\Sigma$ solution in the anti-twist class for $n_N=4$, $n_S=1$, and $p_F=2$ in the range of $y=[y_N,y_S]\approx[0,3.96758]$. The metric functions, $e^V$ (Blue) and $h$ (Orange), are on the left. The scalar fields, $\chi_1$ (Blue), $\chi_2$ (Orange), and $\phi$ (Green) are on the right. Note that $h$ vanishes at the poles.}\label{fig1}}
\end{center}
\end{figure}

We numerically perform the Bekenstein-Hawking entropy integral in section \ref{sec333} and the result matches the Bekenstein-Hawking entropy in \eqref{bhent} with the numerical accuracy of order $10^{-3}$. We present a representative solution in figure \ref{fig1} for the choice in \eqref{ninput} in the range of $y=[y_N,y_S]\approx[0,3.96758]$. The scalar field, $\phi$, takes the values, $\phi|_N\approx0.5025$ and $\phi|_S\approx0.223216$, at the poles. Note that $h$ vanishes at the poles.

There appears to be constraints on the parameter space of $n_{N,S}$, $t_{N,S}$, $p_F$. However, without the analytic expressions of the boundary conditions, it is not easy to specify the constraints.

\section{Gravitational blocks}
\label{sec:gluing}

We now consider the description of our results in terms of gravitational blocks, which may greatly facilitate the potential match with the holographically dual field theories. We first adopt a four-dimensional point of view, elaborating on the basic gravitational building blocks in the presence of abelian charged hypermultiplets. The situation was described in detail in our previous work, \cite{Hristov:2023rel}, and we largely follow the logic there.

A crucial role in the gravitational block description is played by the $(n_V+1)$ $U(1)$ gauge fields $A^I$, $I = 0, 1, .. n_V$ that can be used to gauge the isometries of the hypermultiplet scalar manifold. Via supersymmetry, the gravitini's R-symmetry also becomes charged under a particular linear combination of the gauge fields. After gauging, one generically finds a number of massive vectors $A^m_\alpha = \zeta_{\alpha, I} A^I$ (their mass is generated by a supersymmetry-preserving Higgs mechanism), and the (necessarily massless) R-symmetry vector $A^R = \xi_I A^I$, where the coefficients $\zeta_{\alpha, I}$ and $\xi_I$ take particular constant values on a given background for the hypermultiplet scalars. The choice of $A^m_\alpha$ and $A^R$ is then fixed by the hypermultiplet sector data. The vector multiplet scalar manifold and gauge field kinetic terms are instead defined by the so-called prepotential, $F(X^I)$, a homogeneous function of degree 2 of the sections $X^I$ that determine the complex scalars. We can further define the so-called {\it effective} prepotential $F^\text{eff}$, by setting to zero the sections belonging to the massive vector multiplets,
\be
\label{eq:mainHiggs}
	\zeta_{\alpha, I}\, X^I = 0\ , \quad \forall \alpha\ ,
\ee
as dictated by maximal supersymmetry, obtaining
\be
	F^\text{eff} = F(\zeta_{\alpha, I}\, X^I = 0)\ .
\ee
This is in accordance with the supersymmetry preserving Higgs mechanism, see \cite{Hristov:2010eu, Hosseini:2017fjo}. 

Turning to a more concrete discussion of the gravitational blocks, we start with a single block as defined in \cite{Hosseini:2019iad},
\be
	\cB(X^I, \epsilon) = \frac{i\, \pi}{8\, G_N^{(4)}}\, \frac{F^\text{eff} (X^I)}{\epsilon}\ ,
\ee
where we already used the {\it effective} prepotential and $\epsilon$ can be understood geometrically as an $\Omega$-deformation parameter at each fixed point of the {\it canonical} isometry (generated from the Killing spinor bilinears) on a given background. The on-shell action of a 4d BPS background $M_4$ with positive Euler number $\chi_E (M_4) > 0$ corresponding to the number of fixed points $n_F (M_4)$ is then given by the general {\it gluing} formula
\be
	\cF_{\partial M_4} (\varphi, \epsilon) = \sum_{s = 1}^{n_F (M_4)}\, \sigma_{(s)}\, \cB(X^I_{(s)}, \epsilon_{(s)})\ ,
\ee
where the corresponding identifications of $\epsilon, X^I (\varphi^I, \epsilon)$, and the relative sign $\sigma$ at each different fixed point $s$, together with one overall constraint on the fugacities coming from supersymmetry, are known as a {\it gluing rule} and depend on the particular background. 

The simplest example of Euclidean $AdS_4$ with round $S^3$ boundary exhibits only a single fixed point. The corresponding on-shell action $\cF_{S^3}$ is precisely equal to the free energy of the dual field theory. We find
\be
\label{eq:S3freenergy}
	\cF_{S^3} (\varphi) = \cB(X^I =2 \varphi^I, \epsilon = 1) =  \frac{i\, \pi}{2 G_N^{(4)}}\, F^\text{eff}(\varphi^I)\ ,
\ee
under the constraints, $\xi_I \varphi^I = 2$ and $\zeta_{\alpha, I}\, \varphi^I = 0$. The extremization of the above functional with respect to $\varphi^I$ is the supergravity equivalent of the so-called $F$-maximization, \cite{Jafferis:2010un,Jafferis:2011zi}.

Black holes with spindle horizons correspond to $n_F = 2$ and therefore have two fixed points that are situated precisely at the horizon that we have discussed, at the center of the $AdS_2$ factor and the two conical singularities (or poles) of $\Sigma$. The gluing rules in this case feature the topological numbers of the spindle, as well as the magnetic fluxes of the particular background, as discussed in \cite{Hosseini:2021fge, Faedo:2021nub}. The on-shell action of black holes with spindle horizons (defined by the co-prime integers $n_-, n_+$) is simply given by
\be
\label{eq:1}	
\cF^\sigma_{S^1 \times \Sigma} (\varphi, \epsilon; n_\pm) = \frac{i\, \pi}{8\, G^{(4)}_N\, \epsilon}\,  \left( F^\text{eff} (\varphi^I +\epsilon\, \frak{n}^I) - \sigma F^\text{eff} (\varphi^I -\epsilon\, \frak{n}^I) \right)\ ,
\ee
where $\frak{n}^I$ are the respective magnetic fluxes through the spindle. Supersymmetry further dictates that
\be
\label{eq:fluxes}
	\frak{n}^R = \xi_I \frak{n}^I = \frac{n_++\sigma{n}_-}{n_+n_-}\,, \qquad \frak{n}^m_\alpha = \zeta_{\alpha, I} \frak{n}^I = 0\ ,
\ee
where $\sigma = \pm 1$ reflects the Killing spinor orientation at the two poles of the spindle and is called twist and anti-twist, respectively. We also have the supersymmetric condition
\begin{equation} \label{varphicst}
\varphi^R = \xi_I \varphi^I=\,2 + \frac{n_+-\sigma{n}_-}{n_+n_-}\epsilon\, ,
\end{equation}
together with $\zeta_{\alpha, I} \varphi^I = 0$.

In the absence of electric charges and rotation, the Bekenstein-Hawking entropy is then obtained by extremizing the off-shell entropy function,
\begin{equation}
\label{eq:entropyextrem}
S^\sigma \left(\varphi,\epsilon;\mathfrak{n}_+,\mathfrak{n}_-\right)\,=\, - \frac{i\, \pi}{8\, G^{(4)}_N\, \epsilon}\, \Big( F^\text{eff} \left(\varphi^I +\epsilon\, \frak{n}^I \right)-\sigma\, F^\text{eff} \left(  \varphi^I -\epsilon\, \frak{n}^I \right)\Big)\,,
\end{equation}
where, in order to recover the entropy in terms of the conserved charges, one needs to extremize the above functional with respect to the fugacities $\varphi^I$ and $\epsilon$,
\be
\label{eq:3}
	S^\sigma_\text{on-shell} (q, J; n_\pm ) = S^\sigma (q_I, J, \bar \varphi^I, \bar \epsilon; n_\pm ) \ , \qquad \partial_{\varphi^I, \epsilon}  S^\sigma |_{\bar{\varphi}^I, \bar{\epsilon}} = \partial_{\lambda, \mu^\alpha} S^\sigma |_{\bar{\varphi}^I, \bar{\epsilon} } = 0\ .  
\ee
Note that this is a constrained extremization due to conditions, \eqref{eq:fluxes}-\eqref{varphicst}. Apart from matching the on-shell Bekestein-Hawking entropy of the solutions, we also show that the values of the vector multiplet scalars at the poles of the spindle are precisely related to the extremal values $\bar \varphi^I$ and $\bar \epsilon$. The novel feature of hypermultiplet gauging is that each massive multiplet contributes with an extra constraint, $\zeta_{\alpha, I} \varphi^I = 0$, decreasing the flavor symmetries, or massless $U(1)$ vectors. Again, this can be understood from the supersymmetry-preserving Higgs mechanism that takes place at the poles of the spindle, \cite{Hristov:2010eu,Hosseini:2017fjo}.

\subsection{The class $\cF$ model}

Here we apply the above general procedure to the class $\cF$ model in order to compare with the explicit solution of the previous section. The effective prepotential, see \cite{Hosseini:2020wag}, is given by
\begin{equation}\label{eq:effF}
F^\text{eff} = - \frac{i}{4 m}\, \left( s^2\, X^1 + s^1\, X^2 \right)  \sqrt{X^1 X^2}\ ,
\end{equation} 
and the R-symmetry vector is given by 
\begin{equation}
A^R =3m\, ( A^1 + A^2)\ ,
\end{equation}
which corresponds to $ \xi_1 = \xi_2 = 3 m$. If we further use the parametrization,
\be
	s^{1,2} = - \frac{\tilde \kappa}{6 m}\, \left( 1 \pm \frac{z}{\tilde \kappa} \right)\ ,
\ee
such that
\be
	F^\text{eff} = - \frac{i}{24 m^2}\, \left( z\, (X^1-X^2) - \tilde \kappa\, (X^1+X^2) \right)  \sqrt{X^1 X^2}\ ,
\ee
we find the $AdS_4$ scale of the model to be 
\be
L^2_{AdS_4} = \frac{1}{864 m^4}\, \frac{(3 \tilde \kappa-\sqrt{8 z^2 + \tilde \kappa^2})^2 }{\sqrt{4 z^2+2 \tilde \kappa (\tilde \kappa-\sqrt{8 z^2+\tilde \kappa^2})}}\ \ . 
\ee

We first evaluate the three-sphere free energy in this model, \eqref{eq:S3freenergy}, which gives
\be
	\cF^{\text{class}\cF}_{S^3} (\varphi) = \frac{\pi}{8 m\, G_N^{(4)}}\, \left( s^2\, \varphi^1 + s^1\, \varphi^2 \right)  \sqrt{\varphi^1 \varphi^2}\ ,
\ee 
under the constraint, $\varphi^2 = 2/3m - \varphi^1$. It is easy to extremize the above formula, finding 
\be
	\bar \varphi^1 = \frac{\tilde \kappa+4 z + \sqrt{8 z^2+\tilde \kappa^2}}{12 m z}\ ,
\ee
which corresponds to the superconformal point of the dual theory. We therefore find 
\begin{equation} 
\cF^{\text{class}\cF}_{S^3} := \cF^{\text{class}\cF}_{S^3} (\bar \varphi) = \frac{\pi}{1728 m^4\, G_N^{(4)}}\, \frac{(3 \tilde \kappa-\sqrt{8 z^2 + \tilde \kappa^2} )^2 }{\sqrt{4 z^2+2 \tilde \kappa (\tilde \kappa-\sqrt{8 z^2+\tilde \kappa^2})}}\ ,
\end{equation}
such that
\be
	 \cF^{\text{class}\cF}_{S^3}= \frac{\pi\, L^2_{AdS_4}}{2\, G^{(4)}_N}\, ,
\ee
as expected.

Now we consider the spindle entropy function, given by \eqref{eq:entropyextrem}. The twist condition \eqref{eq:fluxes} becomes
\begin{equation}
\label{eq:constrn}
3 m\, \sum_{i=1}^2\mathfrak{n}^i\,=\,\frac{n_++\sigma{n}_-}{ n_+n_-}\,,
\end{equation} 
and the constraint in \eqref{varphicst} gives
\begin{equation} \label{eq:constrphi}
3m\, \sum_{i=1}^2\varphi^i-\frac{n_+-\sigma{n}_-}{n_+n_-}\epsilon\,=\,2\,.
\end{equation}
We choose $\sigma = -1$ and find for the entropy function,
\be
S^- \left(\varphi,\epsilon;\frak{n}, n_+, n_-\right)\,=- \frac{i\, \pi}{8\, G^{(4)}_N\, \epsilon}\, \Big( F^\text{eff}\left(\varphi^i + \epsilon \frak{n}^I \right) + F^\text{eff}\left(\varphi^i - \epsilon \frak{n}^I \right)  \Big)\ ,
\ee
using \eqref{eq:effF}. Upon extremization, we recover the corresponding black hole entropy,
\be
	S_\text{BH} (\frak{n}, n_+, n_-) = S^- (\bar \varphi, \bar \epsilon; \frak{n}, n_+, n_-)\ ,
\ee
and furthermore the values of the scalar fields at the poles are precisely given by
\be
	4 m\, \frac{\bar \varphi^1 + \frak{n}^1 \bar \epsilon}{s^2 (\bar \varphi^1 + \frak{n}^1 \bar \epsilon) + s^1 (\bar \varphi^2  + \frak{n}^2 \bar \epsilon)} = e^{2 \chi_1 + \chi_2} |_\text{N}\ , \quad 4 m\, \frac{\bar \varphi^1 - \frak{n}^1 \bar \epsilon}{s^2 (\bar \varphi^1 - \frak{n}^1 \bar \epsilon) + s^1 (\bar \varphi^2  - \frak{n}^2 \bar \epsilon)} = e^{2 \chi_1 + \chi_2} |_\text{S}\ , 
\ee
as well as
\be
	4 m\, \frac{\bar \varphi^2 + \frak{n}^2 \bar \epsilon}{s^2 (\bar \varphi^1 + \frak{n}^1 \bar \epsilon) + s^1 (\bar \varphi^2  + \frak{n}^2 \bar \epsilon)} = e^{2 \chi_1 - \chi_2} |_\text{N}\ , \quad 4 m\, \frac{\bar \varphi^2 - \frak{n}^2 \bar \epsilon}{s^2 (\bar \varphi^1 - \frak{n}^1 \bar \epsilon) + s^1 (\bar \varphi^2  - \frak{n}^2 \bar \epsilon)} = e^{2 \chi_1 - \chi_2} |_\text{S}\ , 
\ee
upon the constraints, \eqref{eq:constrn} and \eqref{eq:constrphi}. 

Extremizing the entropy function with the constraints above is however technically challenging and we could not obtain analytic expressions for the entropy with arbitrary choices of $z, n_\pm$. However, any set of specific choices allows for a direct numeric solutions that can be readily compared with the explicit solutions and successfully matched, see the attached {\it .nb} file.

\subsection{Higher-dimensional constructions}

After the original constructions of black hole entropy functions from gravitational blocks in \cite{Hosseini:2019iad} and the localization of the action in terms of nuts and bolts in minimal gauged supergravity in \cite{BenettiGenolini:2019jdz}, there have been various generalizations and complementary approaches in deriving these contributions from the first principles. We will comment on the relation of our results above to two recent approaches based on equivariant localization: the work of Benetti Genolini, Gauntlett and Sparks, \cite{BenettiGenolini:2023kxp}, proposes a direct application of the Berline-Vergne-Atiyah-Bott fixed point formula on the (effective) supergravity action (applied to our construction we can consider either in 4d or in 6d). On the other hand the work of Martelli and Zaffaroni, \cite{Martelli:2023oqk}, starts in 10d/11d and defines the equivariant volume of the (cone over the) manifold that we have compactified upon. These approaches are in agreement whenever applicable at the same time and, while each has some advantages and some limitations when applied in the present context, they can both be used to understand the previous results from fresh perspectives.

\subsubsection{Six-dimensional point of view}

Let us consider how our 4d blocks look like from a perspective of the parent 6d supergravity. Given that the relation between the respective Newton's constants is simply given by
\be
	\frac1{G_N^{(4)}} = \frac{4\pi |1-\frak{g}|}{G_N^{(6)}}\ ,
\ee
we find a single 4d gravitational block to be given in terms of 6d quantities in the following way,
\be
	 \cB_{4d} (X^I, \epsilon) = \frac{\pi^2}{8 m\, G_N^{(6)}}\,  |1-\frak{g}| \frac{ \left( s^2\, X^1 + s^1\, X^2 \right)  \sqrt{X^1 X^2}}{\epsilon}\ .
\ee
Next we note that a single 6d building block is instead based on the function,
\be
\label{eq:6dblock}
	F_{6d} = (X^1 X^2)^{3/2}\ ,
\ee
which was introduced already in \cite{Hosseini:2019iad} and has been tested successfully on backgrounds exhibiting fixed points of the canonical isometry, see \cite{Hosseini:2018usu}. The formulae presented in \cite{Hosseini:2019iad} however did not take into account the possibility of fixed two-submanifolds, or bolts, which also arise naturally in supersymmetric solutions. The Riemann surface of genus $\frak{g}$ is a prime example of a bolt-type geometry and this situation is more naturally described by the approach in \cite{BenettiGenolini:2023kxp}. Although the latter reference only considered minimal 6d supergravity, the approach is conceptually applicable to the matter-coupled case here as well. In this sense it is interesting to rewrite our findings in what we believe is the natural 6d form,
\be
	\cB_{6d\, \text{on}\, \Sigma_\frak{g}}	:= \cB_{4d} (X^I, \epsilon) =  \frac{\pi^2}{12 m\, \epsilon\, G_N^{(6)}}\,  |1-\frak{g}|\, \sum_I s^I\, \frac{\partial F_{6d}}{\partial X^I}\ ,
\ee
where the effect of the Riemann surface can be seen both in the overall factor of $(1 - \frak{g})$ and in the contribution from the magnetic fluxes $s^{1,2}$. The explicit form of $F_{6d}$ should instead be considered as uniquely fixed by the choice of matter-coupled theory, as also discussed in \cite{Hosseini:2018usu}.

\subsubsection{Ten-dimensional point of view}

We now go to the grand-parent 10d supergravity (here we commit ourselves to the massive IIA uplift) and briefly sketch the logic in \cite{Martelli:2023oqk,Colombo:2023fhu}.\footnote{We would like to thank Alberto Zaffaroni for explanations on this.} From this perspective we are able to justify the explicit form of the 6d potential $F_{6d}$ dual to the five-sphere partition function from a geometric perspective. More explicitly, it can be related to the equivariant volume of $\mathbb{C}^2$ which was argued in \cite{Martelli:2023oqk} to be the relevant quantity describing the internal space in the uplift from 6d to 10d, a hemisphere S$^4$. The equivariant volume of  $\mathbb{C}^2$, in the presence of zero-dimensional ``higher time'' constant $c$ (the dual of a flux over the full hemisphere), is
\be
	\mathbb{V}_{\mathbb{C}^2} = \frac{e^{ c -\lambda_I X^I}}{X^1 X^2}\ ,
\ee
where $X^{1,2}$ play the role of equivariant parameters, the $\lambda_{1,2}$ provide a parametrization of the K\"ahler moduli that do not play any role here (they have a crucial role for black hole and spindle solutions).
The $n-$th order term in the Taylor expansion of the equivariant volume is defined as
\be
	\mathbb{V}_{\mathbb{C}^2}^{(n)} \propto \frac{(c-\lambda_I X^I)^n}{ X^1 X^2}\ .
\ee
The authors of \cite{Martelli:2023oqk, Colombo:2023fhu} then define an extremization principle fixing (up to an irrelevant gauge freedom) the constants $c, \lambda_I$,~\footnote{Here we are only sketching the calculation and for our purposes have rescaled the parameters $c, \lambda_I$ with respect to the original references in order not to carry powers of the $N$, the number of D4-branes and dual gauge group rank.}
\be
	\frac{\partial \mathbb{V}^{(3)}}{\partial c} \propto 1\ , \quad \Rightarrow \quad (c-\lambda_I X^I) \propto (X^1 X^2)^{1/2}\ .
\ee
Finally, we find
\be
	F_{6d} \propto \mathbb{V}_{\mathbb{C}^2}^{(5)} \propto (X^1 X^2)^{3/2}\ ,
\ee
in agreement with our previous results, \eqref{eq:6dblock}.  

\section{Conclusions} \label{sec:5}

In this study, we have derived new $AdS_2\times\Sigma$ solutions which serve as dual descriptions for three-dimensional field theories belonging to class $\mathcal{F}$. These solutions were constructed within the framework of $\mathcal{N}=2$ gauged supergravity featuring vector multiplets and a hypermultiplet obtained from the consistent truncation of $F(4)$ gauged supergravity on a Riemann surface. Through numerical methods, we obtained solutions and calculated the Bekenstein-Hawking entropy for postulated black hole configurations, finding precise agreement with the results derived from gravitational blocks.

From six-dimensional point of view, the $AdS_2\times\Sigma\times\Sigma_\mathfrak{g}$ solutions that we presented can be regarded as an extension of previously studied $AdS_2\times\Sigma_{\mathfrak{g}_1}\times\Sigma_{\mathfrak{g}_2}$ solutions, \cite{Suh:2018szn, Hosseini:2018usu, Suh:2018tul}, where $\Sigma_\mathfrak{g}$ denotes a Riemann surface of genus $\mathfrak{g}$. For the latter solutions, their Bekenstein-Hawking entropy was previously computed via microscopic counting using the topologically twisted index of five-dimensional SCFTs on $S^1\times\Sigma_{\mathfrak{g}_1}\times\Sigma_{\mathfrak{g}_2}$, see \cite{Hosseini:2018uzp, Crichigno:2018adf}. Recent investigations have defined the spindle index for three-dimensional SCFTs, \cite{Inglese:2023wky, Inglese:2023tyc, Colombo:2024mts, Mauch:2024uyt}, and explored it from the perspective of gravity duals, \cite{Boido:2023ojv}. It would be particularly intriguing to compute the large $N$ spindle index of the class $\mathcal{F}$ theories discussed herein, or alternatively, to compute the partition function of five-dimensional SCFTs on $S^1\times\Sigma\times\Sigma_\mathfrak{g}$ and verify its precise reproduction of the Bekenstein-Hawking entropy.

Our reformulation of the entropy function in terms of gravitational blocks is expected to greatly facilitate the matching of field theory results. In this context, we also related our findings with several parallel developments concerning spindle black holes, \cite{Boido:2022mbe, Boido:2022iye}, including the utilization of GK geometry, \cite{Kim:2006qu, Gauntlett:2007ts}, as well as recent investigations employing equivariant localization, \cite{BenettiGenolini:2023kxp, Martelli:2023oqk, BenettiGenolini:2023yfe, BenettiGenolini:2023ndb, Colombo:2023fhu, BenettiGenolini:2024kyy}. All of these constructions based on gravitational blocks make our holographically dual prediction very natural. However, it is still important to provide the explicit solutions as done here, in order to ensure the lack of any local or global obstructions.

\bigskip

\leftline{\bf Acknowledgements}
\noindent We wish to thank Seyed Morteza Hosseini and Alberto Zaffaroni for illuminating discussions. The study of KH is financed by the European Union- NextGenerationEU, through the National Recovery and Resilience Plan of the Republic of Bulgaria, project No BG-RRP-2.004-0008-C01. KH wishes to thank the kind hospitality of the MATRIX Program, “New Deformations of Quantum Field and Gravity Theories,” during the final stage of this project. MS was supported by the Kumoh National Institute of Technology. 


\appendix
\section{$F(4)$ gauged supergravity on a Riemann surface} \label{appA}
\renewcommand{\theequation}{A.\arabic{equation}}
\setcounter{equation}{0} 

\subsection{The $\mathcal{N}=2$ formalism} \label{appA11}

The consistent truncation of $F(4)$ gauged supergravity on a Riemann surface consists of the gravity multiplet with the graviton and graviphoton, $\{g_{\mu\nu}, A^0_\mu\}$, two vector multiplets with a vector and complex scalar, $\{A^i_\mu, \chi_i\}$, and one hypermultiplet (known as the {\it universal} hypermultiplet) with four real scalars, $\{\phi, \sigma, \xi^0, \tilde{\xi}_0\}$. The scalar fields from the vector multiplets and the hypermultiplet parametrize the coset manifolds,
\begin{equation}
\mathcal{M}_v\times\mathcal{M}_h\,=\,\left(\frac{SU(1,1)}{U(1)}\right)^2\,\times\,\frac{SU(2,1)}{S(U(2)\,\times\,U(1))}\,,
\end{equation}
which is a product of special K\"ahler and quaternionic manifolds, respectively. 

The following more detailed discussion is in close parallel to appendix A of \cite{Hristov:2023rel}, since the scalar model shares many similarities.

{\bf Universal hypermultiplet:} Here we gather the relevant quantities and specific gaugings of the universal hypermultiplet, given by the metric $\mathcal{M}_h$. Written in terms of real coordinates, $\{\phi,\,\sigma,\,\xi^0,\,\tilde{\xi}_0\}$, the metric is  
\begingroup
\renewcommand*{\arraystretch}{1.2}
\begin{equation}
h = 
\begin{pmatrix}
1 & 0 & 0 & 0 \\
0 & \frac{1}{4} e^{4\phi} & - \frac{1}{8} e^{4\phi} \tilde{\xi}_0 & \frac{1}{8} e^{4\phi} \xi^0 \\
0 & - \frac{1}{8} e^{4\phi} \tilde{\xi}_0\, & \, \frac{1}{4} e^{2\phi}\left(1 + \frac{1}{4} e^{2\phi}(\tilde{\xi}_0)^2\right) & -\frac{1}{16} e^{4\phi} \xi^0 \tilde{\xi}_0 \\
0 &  \frac{1}{8} e^{4\phi} \xi^0 & -\frac{1}{16} e^{4\phi} \xi^0 \tilde{\xi}_0 &  \frac{1}{4} e^{2\phi}\left(1 + \frac{1}{4} e^{2\phi} \left(\xi^0\right)^2\right)
\end{pmatrix}.
\end{equation}
\endgroup
The isometry group, $SU(2,1)$, has eight generators; two of these are used for gauging in the model we consider explicitly below, generating the group, $\mathbb{R}\times{U}(1)$.~\footnote{See e.g.\ \cite{Halmagyi:2011xh} for a careful discussion of the isometries and the physical outcome of their gauging.} The corresponding Killing vectors are
\begin{equation}
k^\mathbb{R}\,= \,\partial_\sigma\,, \qquad k^{U(1)}\,=\,-\tilde{\xi}_0\partial_{\xi^0}+\xi^0\partial_{\tilde{\xi}_0}\ .
\end{equation}
These two isometries are gauged by a particular linear combination of the vector fields in the theory. One defines Killing vectors with a symplectic index corresponding to each of the full set of electric and magnetic gauge fields at our disposal. The moment maps associated to these two Killing vectors are
\begin{equation}
P^\mathbb{R}\,=\,\left(0,\,0,\,-\frac{1}{2}e^{2\phi}\right), \qquad P^{U(1)}\,=\,\left(\tilde{\xi}_0e^\phi,\,-\xi^0e^\phi,\,1-\frac{1}{4}\left((\xi^0)^2+(\tilde{\xi}_0)^2\right)e^{2\phi}\right)\ . 
\end{equation}

In order to make sure the moment maps are strictly in the third direction, we can set $\xi^0\,=\,\tilde{\xi}_0\,=\,0$ guaranteeing that $k^{U(1)} = 0$ independent of the details of the scalar manifold for the vector multiplets. On the contrary, $k^\mathbb{R}\neq0$ always and thus we would find a genuine constraint on the vector multiplets from this type of gauging. Now the moment map $P$ remains non-zero only along the third direction, 
\begin{equation}
P^\mathbb{R}\,=\,\left(0,\,0,\,-\frac{1}{2}e^{2\phi}\right)\,, \qquad P^{U(1)}(\xi^0\,=\,\tilde{\xi}_0\,=\,0)\,=\,\left(0,\,0,\,1\right)\,.
\end{equation}
From now on we will only discuss this third, or $z$, component of the moment maps. The choice of setting  $\xi^0\,=\,\tilde{\xi}_0\,=\,0$ is not in itself a subtruncation to a smaller $\cN=2$ supergravity, but it can always be made on a given background without breaking further supersymmetry.

Note also that this way of solving the hyperscalar equations for the universal hypermultiplet, by setting $\xi^0=\tilde{\xi}_0=0$ and keeping only $k^\sigma$ non-vanishing also means that the $SU(2)$ connection takes a simple form, 
\begin{equation}
\omega^x\,=\,e^\phi{\rm d}\xi^0\,=\,0\,, \qquad \omega^y\,=\,e^\phi{\rm d}\tilde{\xi}_0\,=\,0\,, \qquad \omega^z\,=\,-\frac{1}{2}e^{2\phi}\left({\rm d}\sigma+\xi^0{\rm d}\tilde{\xi}_0\right)\,=\,-\frac{1}{2}e^{2\phi}{\rm d}\sigma\,.
\end{equation}
The relevant part of the corresponding $SU(2)$ curvature, defined as $\Omega^x := {\rm d} \omega^x - \tfrac12 \epsilon^{x y z} \omega^y \wedge \omega^z$, is therefore
\be
	\Omega^z_{\sigma \phi} = - \Omega^z_{\phi \sigma} =\frac12 e^{2\phi}\, .
\ee

{\bf Vector multiplets:} The full model is further specified by the vector multiplet geometry, which is given by the so-called ST model, $i.e.$, the coset space $\mathcal{M}_v$ above with a prepotential given by 
\begin{equation}\label{F(4)}
	F =  - i\, X^0 \sqrt{X^1 X^2}\ ,
\end{equation}
that defines (see below) the corresponding metric. In order to simplify the model from the start, we assume that there are no axions, such that the two complex scalar fields, $\chi_1$ and $\chi_2$, are real. The Lagrangian and supersymmetry variations then follow from the choice of holomorphic sections,
\begin{equation}
X^0\,=\,1\,, \qquad X^1\,=\,e^{2\chi_1+\chi_2}\,, \qquad X^2\,=\,e^{2\chi_1-\chi_2}\,,
\end{equation}
leading to a K\"ahler potential,
\begin{equation}
e^{-K}\,=\,\left(e^{2\chi_1}+e^{2\bar{\chi}_1}\right)\left(1+\cosh\left(\chi_2-\bar\chi_2\right)\right)\,=\,4e^{2\chi_1}\,,
\end{equation}
with quantities,
\begin{equation}
K_{\chi_1}\,=\,-1\,, \qquad K_{\chi_2}\,=\,0\,, \qquad g_{\chi_1\bar{\chi}_2}\,=\,2g_{\chi_2\bar\chi_2}\,=\,1\,.
\end{equation}
The so-called period matrix, which defines the gauge field couplings, is given by
\begin{equation}
\mathcal{N}\,=\,-\frac{i}{2}\,\text{diag}\left(2e^{2\chi_1},e^{-2\chi_1-2\chi_2},e^{-2\chi_1+2\chi_2}\right)\,,
\end{equation}
such that $\text{Re}\mathcal{N}=0$. 

{\bf Gauging:} The consistent truncation with universal hypermultiplet gauging coming from the compactification of $F(4)$ gauged supergravity on a Riemann surface, features a mixed dyonic gaugings. We have a hypermultiplet gauging,
\begin{equation}
k_I\,=\,\left(-4mk^\mathbb{R},\,3mk^{U(1)}+s_2k^\mathbb{R},\,3mk^{U(1)}+s_1k^\mathbb{R}\right)\,,
\end{equation}
with $m$ the six-dimensional gauge coupling and the twisting parameters, $s_1$ and $s_2$,
\begin{equation}
s_1+s_2\,=\,-\frac{\tilde{\kappa}}{3m}\,,
\end{equation}
where $\tilde{\kappa}=\pm1$ is the curvature of the Riemann surface. Thus we have
\begin{equation}
P_I^z\,=\,\left(2me^{2\phi},\,3m-\frac{1}{2}s_2e^{2\phi},\,3m-\frac{1}{2}s_1e^{2\phi}\right)\,.
\end{equation}

{\bf Lagrangian and supersymmetry variations:} Following the conventions of \cite{Andrianopoli:1996cm}, the Lagrangian, after the simplifications of taking real vector multiplet scalars and setting $\xi^0\,=\,\tilde{\xi}_0\,=\,0$, reads~{\footnote{Only in this appendix \ref{appA11}, we employ the mostly minus signature and stick to the notation and conventions in \cite{Andrianopoli:1996cm}.}}
\begin{equation}
e^{-1}\mathcal{L}\,=\,\frac{1}{2}R+\left(\partial\chi_1\right)^2+\frac{1}{2}\left(\partial\chi_2\right)^2+\left(\partial\phi\right)^2+\frac{1}{4}e^{4\phi}\left(D \sigma\right)^2+\frac{1}{4}\text{Im}\mathcal{N}_{IJ}F^I_{\mu\nu}F^{I\mu\nu}-\mathcal{V}\,,
\end{equation}
with the gauge covariant derivative,
\begin{equation}
\label{eq:Dsigma}
D_\mu\sigma\,=\,\partial_\mu\sigma+A^m_\mu\,,
\end{equation}
where we define the massive vector,
\begin{equation}
A^m\,=\,-4mA^0+s_2A^1+s_1A^2\,,
\end{equation}
and $I,J=0,1,2$. The scalar potential follows straightforwardly from the data given above, and is discussed further below. The R-symmetry gauge field is essentially chosen by the orientation of the Killing vector $k^{U(1)}$,
\begin{equation}
A^R\,=\,3m\left(A^1+A^2\right)\,.
\end{equation}
The supersymmetry variations of gravitino, gaugino and hyperino are given by, respectively,\footnote{In order to bring the hyperino variation to exhibit a free $SU(2)$ index, we make use of the identity $U_{u \alpha A} U_v^{\alpha B} = \tfrac12 h_{u v} \delta_A{}^B + \tfrac{i}2 \Omega^x_{u v} \sigma^3{}_A{}^B$, see \cite{Ceresole:2001wi}.}
\begin{align} \label{gravitino0}
\delta\psi_{\mu{A}}\,=&\,\nabla_\mu\varepsilon_A+\frac{i}{2}\left(4\sqrt{2}A^0_\mu-\frac{1}{2}e^{2\phi}\nabla_\mu\sigma\right)\sigma^3{}_A{}^B\varepsilon_B-\frac{1}{2}e^{K/2}P_I^3X^I\gamma_\mu\sigma^3{}_{AB}\varepsilon^B \notag \\ 
&+2ie^{K/2}X^I\text{Im}\mathcal{N}_{IJ}F^{J}_{\mu\nu}\gamma^\nu\epsilon_{AB}\varepsilon^B\,, \\ \label{gaugino0}
\delta\lambda^{iA}\,=&\,i\partial_\mu{z}^i\gamma^\mu\varepsilon^A+ie^{K/2}g^{ij}\left(\partial_j+K_j\right)X^IP^3_I\sigma^{3AB}\varepsilon_B \notag \\ 
&-e^{K/2}g^{ij}\left(\partial_j+K_j\right)X^I\text{Im}\mathcal{N}_{IJ}F^{J}_{\mu\nu}\gamma^{\mu\nu}\epsilon^{AB}\varepsilon_B\,, \\ \label{hyperino0}
\delta\zeta_\alpha\,=&\,U_{u\alpha{A}}\left(i\nabla_\mu{q}^u\gamma^\mu\varepsilon^A+2e^{K/2}k_I^uX^I\epsilon^{AB}\varepsilon_B\right)\,,
\end{align}
and we define quantities for later convenience,
\begin{align}
W\,\equiv&\,e^{K/2}P_I^3X^I\,, \notag \\
H_{\mu\nu}\,\equiv&\,e^{K/2}X^I\text{Im}\mathcal{N}_{IJ}F^{J}_{\mu\nu}\,.
\end{align}

\subsection{Parametrizations and equations of motion} \label{appA11}

The bosonic Lagrangian is, \cite{Hosseini:2020wag},
\begin{align} \label{lagapp}
e^{-1}\mathcal{L}\,&=\,\frac{1}{2}R-\partial_\mu\chi_1\partial^\mu\chi_1-\frac{1}{2}\partial_\mu\chi_2\partial^\mu\chi_2-\partial_\mu\phi\partial^\mu\phi-\frac{1}{4}e^{4\phi}D_\mu\sigma{}D^\mu\sigma-\mathcal{V} \notag \\
&-\frac{1}{8}\Big[2e^{2\chi_1}F^0_{\mu\nu}F^{0\mu\nu}+e^{-2\chi_1-2\chi_2}F^1_{\mu\nu}F^{1\mu\nu}+e^{-2\chi_1+2\chi_2}F^2_{\mu\nu}F^{2\mu\nu}\Big]\,,
\end{align}
where we have
\begin{equation}
D\sigma\,=\,d\sigma-4mA^0+s_2A^1+s_1A^2\,.
\end{equation}
The scalar potential,
\begin{align}
\mathcal{V}\,=\,&-3e^{2\phi}\left(8m^2\cosh\chi_2-m\left(s_1+s_2\right)e^{2\chi_1}\right)+\frac{1}{4}e^{4\phi}\left(8m^2e^{-2\chi_1}+s_1^2e^{2\chi_1-2\chi_2}+s_2^2e^{2\chi_1+2\chi_2}\right) \notag \\
&-18m^2e^{2\chi_1}\,,
\end{align}
can be given by
\begin{equation}
\mathcal{V}\,=\,\left(\frac{\partial\,W}{\partial\,\chi_1}\right)^2+2\left(\frac{\partial\,W}{\partial\,\chi_2}\right)^2+\left(\frac{\partial\,W}{\partial\phi}\right)^2-3W^2\,,
\end{equation}
and the superpotential is
\begin{equation}
W\,=\,3me^{\chi_1}\cosh\chi_2+\frac{1}{4}e^{2\phi}\left(4me^{-\chi_1}-s_1e^{\chi_1-\chi_2}-s_2e^{\chi_1+\chi_2}\right)\,.
\end{equation}

The supersymmetry variations of gravitino, gaugino and hyperino are given by
\begin{align}
\delta\psi_{\mu\,A}\,\sim\,&\,2\nabla_\mu\varepsilon_A-iB_\mu\sigma^3{}_A{}^B\varepsilon_B-W\gamma_\mu\sigma^3{}_{AB}\varepsilon^B+4iH_{\mu\nu}\gamma^\nu\epsilon_{AB}\varepsilon^B\,, \notag \\
\delta\lambda^{iA}\,\sim\,&\,\partial_\mu{}u_i\gamma^\mu\varepsilon^A+\partial_{u_i}W\sigma^{3AB}\varepsilon_B-i\partial_{u_i}H_{\mu\nu}\gamma^{\mu\nu}\epsilon^{AB}\varepsilon_B\,, \notag \\
U^{\alpha\, A}_\phi\, \delta\zeta_\alpha\,\sim\, &  U^{\alpha\, A}_\sigma\, \delta\zeta_\alpha\,\sim\,\,\partial_\mu\phi\gamma^\mu\varepsilon^A+\frac{i}{2}\partial_\phi{B}_\mu\gamma^\mu \sigma^3{}_B{}^A\varepsilon^B+\partial_\phi\,W\sigma^{3AB}\varepsilon_B\,,
\end{align}
where we have
\begin{align}
H_{\mu\nu}\,=&\,-\frac{1}{2}\left(2e^{\chi_1}F^{0}_{\mu\nu}+e^{-\chi_1-\chi_2}F^{1}_{\mu\nu}+e^{-\chi_1+\chi_2}F^{2}_{\mu\nu}\right)\,, \notag \\
B_\mu\,=&\,-3m\left(A^1_\mu+A^2_\mu\right)+\frac{1}{2}e^{2\phi}D_\mu\sigma\,.
\end{align}
We introduce $F^{IJ}$ from $F^{I}$,
\begin{align}
F^{0}\,=&\,e^{-\chi_1}\left(\overline{F}^{12}+\overline{F}^{56}\right)\,, \notag \\
F^{1}\,=&\,e^{\chi_1+\chi_2}\left(-\overline{F}^{12}+2\overline{F}^{34}+\overline{F}^{56}\right)\,, \notag \\
F^{2}\,=&\,e^{\chi_1-\chi_2}\left(-\overline{F}^{12}-2\overline{F}^{34}+\overline{F}^{56}\right)\,,
\end{align}
and define them by
\begin{equation}
\overline{F}_{\mu\nu}^{12}\,=\,-4\partial_{\chi_1}H_{\mu\nu}\,, \qquad
\overline{F}_{\mu\nu}^{34}\,=\,-4\partial_{\chi_2}H_{\mu\nu}\,, \qquad
\overline{F}_{\mu\nu}^{56}\,=\,4H_{\mu\nu}\,.
\end{equation}

We introduce a complex Dirac spinor, $\epsilon$, from the Weyl spinors, $\epsilon^A$,
\begin{equation}
\epsilon\,=\,\varepsilon_1+\varepsilon^2\,.
\end{equation}
The supersymmetry variations reduce to
\begin{align}
\Big[2\nabla_\mu-iB_\mu-W\gamma_\mu+4iH_{\mu\nu}\gamma^\nu\Big]\epsilon\,=&\,0\,, \notag \\
\Big[\partial_\mu\chi_1\gamma^\mu+\partial_{\chi_1}W+i\partial_{\chi_1}H_{\mu\nu}\gamma^{\mu\nu}\Big]\epsilon\,=&\,0\,, \notag \\
\Big[\frac{1}{2}\partial_\mu\chi_2\gamma^\mu+\partial_{\chi_2}W+i\partial_{\chi_2}H_{\mu\nu}\gamma^{\mu\nu}\Big]\epsilon\,=&\,0\,, \notag \\
\left[\partial_\mu\phi\gamma^\mu+\partial_\phi{W}+\frac{i}{2}\partial_\phi{B}_\mu\gamma^\mu\right]\epsilon\,=&\,0\,.
\end{align}

We present the equations of motion from the Lagrangian in \eqref{lagapp}. The Einstein equations are
\begin{align}
R_{\mu\nu}-&\frac{1}{2}Rg_{\mu\nu}+g^2\mathcal{V}g_{\mu\nu}-2\left(T_{\mu\nu}^\phi+T_{\mu\nu}^{\chi_1}+2T_{\mu\nu}^{\chi_2}\right)-\frac{1}{2}e^{4\phi}T_{\mu\nu}^\sigma \notag \\
-&2e^{2\chi_1}T_{\mu\nu}^{A^0}-e^{-2\chi_1-2\chi_2}T_{\mu\nu}^{A^1}-e^{-2\chi_1+2\chi_2}T_{\mu\nu}^{A^2}\,=\,0\,,
\end{align}
where the energy-momentum tensors are
\begin{align}
T_{\mu\nu}^X\,=&\,\partial_\mu{X}\partial_\nu{X}-\frac{1}{2}g_{\mu\nu}\partial_\rho{X}\partial^\rho{X}\,, \notag \\
T_{\mu\nu}^{A^I}\,=&\,g^{\rho\sigma}F_{\mu\rho}^I{F}_{\nu\sigma}^I-\frac{1}{4}g_{\mu\nu}F_{\rho\sigma}^I{F}^{I\rho\sigma}\,,
\end{align}
and $X$ denotes a scalar field. The Maxwell equations are
\begin{align}
\partial_\nu\left(\sqrt{-g}e^{2\chi_1}F^{0\mu\nu}\right)-m\sqrt{-g}e^{4\phi}g^{\mu\nu}D_\nu\sigma\,=&\,0\,, \notag \\
\partial_\nu\left(\sqrt{-g}e^{-2\chi_1-2\chi_2}F^{1\mu\nu}\right)+\frac{s_2}{2}\sqrt{-g}e^{4\phi}g^{\mu\nu}D_\nu\sigma\,=&\,0\,, \notag \\
\partial_\nu\left(\sqrt{-g}e^{-2\chi_1+2\chi_2}F^{2\mu\nu}\right)+\frac{s_1}{2}\sqrt{-g}e^{4\phi}g^{\mu\nu}D_\nu\sigma\,=&\,0\,.
\end{align}
The scalar field equations are
\begin{align}
&\frac{1}{\sqrt{-g}}\partial_\mu\left(\sqrt{-g}g^{\mu\nu}\partial_\nu\chi_1\right)-\frac{1}{2}\frac{\partial\mathcal{V}}{\partial\chi_1} \notag \\
&-\frac{1}{4}\left(2e^{2\chi_1}F_{\mu\nu}^0F^{0\mu\nu}-e^{-2\chi_1-2\chi_2}F_{\mu\nu}^1F^{1\mu\nu}-e^{-2\chi_1+2\chi_2}F_{\mu\nu}^2F^{2\mu\nu}\right)\,=\,0\,, \notag \\ \notag \\
&\frac{1}{2}\frac{1}{\sqrt{-g}}\partial_\mu\left(\sqrt{-g}g^{\mu\nu}\partial_\nu\chi_2\right)-\frac{1}{2}\frac{\partial\mathcal{V}}{\partial\chi_2} \notag \\
&+\frac{1}{4}\left(e^{-2\chi_1-2\chi_2}F_{\mu\nu}^2F^{2\mu\nu}-e^{-2\chi_1+2\chi_2}F_{\mu\nu}^3F^{3\mu\nu}\right)\,=\,0\,,
\end{align}
and
\begin{equation}
\frac{1}{\sqrt{-g}}\partial_\mu\left(\sqrt{-g}g^{\mu\nu}\partial_\nu\phi\right)-\frac{1}{2}\frac{\partial\mathcal{V}}{\partial\phi}-\frac{1}{2}e^{4\phi}D_\mu\sigma{D}^\mu\sigma\,=\,0\,.
\end{equation}

\subsection{Truncation to minimal gauged supergravity} \label{appA1}

There is a truncation to minimal gauged supergravity. We set the scalar fields to be at their values of the $AdS_4$ vacuum, \eqref{radii}, and impose the gauge fields to be
\begin{align}
A\,\equiv\,e^{\chi_{1*}}A^0\,=\,e^{-\chi_{1*}-\chi_{2*}}A^1\,=\,e^{-\chi_{1*}+\chi_{2*}}A^2\,,
\end{align}
where $\chi_{i*}$ are the values of the scalar fields at the vacuum, \eqref{potads4}. Then we find
\begin{equation}
e^{-1}\mathcal{L}\,=\,\frac{1}{2}R-\mathcal{V}_*-\frac{1}{2}F_{\mu\nu}F^{\mu\nu}\,,
\end{equation}
where $F=dA$ and $\mathcal{V}_*$ is the value of the scalar potential at the vacuum, \eqref{radii}. It is the action of minimal gauged supergravity with the normalization employed in \cite{Ferrero:2020twa}.

\section{The BPS equations} \label{appB}
\renewcommand{\theequation}{B.\arabic{equation}}

We consider the metric and gauge fields, 
\begin{align}
ds^2\,=&\,e^{2V}ds_{AdS_2}^2+f^2dy^2+h^2dz^2\,, \notag \\
A^I\,=&\,a^Idz\,,
\end{align}
where $V$, $f$, $h$, and $a^I$, $I=0,1,2$, depends only on the coordinate, $y$. We introduce the gamma matrices,
\begin{equation}
\gamma^m\,=\,\Gamma^m\otimes\sigma^3\,, \qquad \gamma^2\,=\,1_2\otimes\sigma^1\,, \qquad \gamma^3\,=\,1_2\otimes\sigma^2\,,
\end{equation}
and the spinors,
\begin{equation}
\epsilon\,=\,\psi\otimes\chi\,,
\end{equation}
where $\Gamma^m$ are two-dimensional gamma matrices. The two-dimensional spinor satisfies
\begin{equation}
D_m\psi\,=\,\frac{1}{2}\kappa\Gamma_m\psi\,,
\end{equation}
where $\kappa=\pm1$.

{\bf Summary:} As it parallels the derivation given in appendix B of \cite{Hristov:2023rel}, we do not present the derivation of the BPS equations and present the summary of BPS equations. 

A solution of the projection condition is
\begin{equation} \label{intxi}
\epsilon\,=\,e^{i\frac{\xi}{2}\gamma^3}\eta\,,
\end{equation}
where we introduced an angular parameter, $\xi$. For $\xi\ne0$, we find the complete BPS equations,
\begin{align}
f^{-1}\xi'\,=&\,2W\cos\xi+\kappa{e}^{-V}\,, \notag \\
f^{-1}V'\,=&\,W\sin\xi\,, \notag \\
f^{-1}\chi_1'\,=&\,-\partial_{\chi_1}W\sin\xi\,, \notag \\
f^{-1}\chi_2'\,=&\,-2\partial_{\chi_2}W\sin\xi\,, \notag \\
f^{-1}\phi'\,=&\,-\frac{\partial_\phi{W}}{\sin\xi}\,, \notag \\
f^{-1}\frac{h'}{h}\sin\xi\,=&\,\kappa{e}^{-V}\cos\xi+W\left(1+\cos^2\xi\right)\,,
\end{align}
with two constraints,
\begin{align}
\left(s-B_z\right)\sin\xi\,=&\,-2Wh\cos\xi-\kappa{h}e^{-V}\,, \notag \\
\partial_\phi{W}\cos\xi\,=&\,\frac{1}{2}\partial_\phi{B}_z\sin\xi{h}^{-1}\,.
\end{align}
The field strengths of gauge fields are given by
\begin{align}
\partial_{\chi_1}H_{23}\,=&\,-\frac{1}{4}\partial_{\chi_1}W\cos\xi\,, \notag \\
\partial_{\chi_2}H_{23}\,=&\,-\frac{1}{4}\partial_{\chi_2}W\cos\xi\,, \notag \\
H_{23}\,=&\,-\frac{1}{4}W\cos\xi-\frac{1}{4}\kappa{e}^{-V}\,.
\end{align}
The BPS equations are consistent with the equations of motion from the Lagrangian.

There is also a relation,
\begin{equation}
\partial_yW\,=\,-f\sin\xi\left[\left(\partial_{\chi_1}W\right)^2+2\left(\partial_{\chi_2}W\right)^2+\frac{1}{\sin^2\xi}\left(\partial_\phi{W}\right)^2\right]\,,
\end{equation}
and the superpotential is monotonic along the BPS flow if the sign of $f\sin\xi$ is not changing.

There is an integral of the BPS equations,
\begin{equation}
he^{-V}\,=\,k\sin\xi\,,
\end{equation}
where $k$ is a constant. We eliminate $h$ with the integral of motion and obtain
\begin{align}
f^{-1}\xi'\,=&\,-k^{-1}\left(s-B_z\right)e^{-V}\,, \notag \\
f^{-1}V'\,=&\,W\sin\xi\,, \notag \\
f^{-1}\chi_1'\,=&\,-\partial_{\chi_1}W\sin\xi\,, \notag \\
f^{-1}\chi_2'\,=&\,-2\partial_{\chi_2}W\sin\xi\,, \notag \\
f^{-1}\phi'\,=&\,-\frac{\partial_\phi{W}}{\sin\xi}\,,
\end{align}
with two constraints,
\begin{align} \label{constraint22}
\left(s-B_z\right)\,=&\,-k\left(2We^V\cos\xi+\kappa\right)\,, \notag \\
\partial_\phi{W}\cos\xi\,=&\,\frac{1}{2}k^{-1}e^{-V}\partial_\phi{B}_z\,.
\end{align}

From the definition of $B_z$, we obtain
\begin{equation}
\partial_\phi{B}_z\,=\,e^{2\phi}D_z\sigma\,.
\end{equation}
For $\phi\ne0$, from the second equation in \eqref{constraint22}, we obtain
\begin{equation} \label{dzpsi}
D_z\sigma\,=\,\frac{2ke^V\partial_\phi{W}\cos\xi}{e^{2\phi}}\,,
\end{equation}
and the right hand side is independent of $\phi$. By differentiating \eqref{dzpsi}, we obtain fluxes expressed by
\begin{equation}
F^I_{yz}\,=\,\left(a^I\right)'\,=\,\left(\mathcal{I}^{(I)}\right)'\,,
\end{equation}
where we introduce
\begin{align}
\mathcal{I}^{(0)}\,\equiv&\,\frac{1}{\sqrt{2}}ke^V\cos\xi{e}^{-\chi_1}\,, \notag \\
\mathcal{I}^{(1)}\,\equiv&\,\frac{1}{\sqrt{2}}ke^V\cos\xi{e}^{\chi_1+\chi_2}\,, \notag \\
\mathcal{I}^{(2)}\,\equiv&\,\frac{1}{\sqrt{2}}ke^V\cos\xi{e}^{\chi_1-\chi_2}\,.
\end{align}

A symmetry of the BPS equations is
\begin{equation} \label{hsymm}
h\,\rightarrow\,-h\,, \qquad z\,\rightarrow\,-z\,,
\end{equation}
if $B_z\rightarrow-B_z$, $s\rightarrow-s$, $a^I\rightarrow-a^I$, $k\rightarrow-k$ and $F_{23}^I\rightarrow-F_{23}^I$. The frame is invariant under the transformation. By this symmetry we fix $h\le0$ in the main text.

\bibliographystyle{JHEP}
\bibliography{20231102.bib}

\end{document}